\documentclass[letterpaper]{JHEP3}
\usepackage[centertags]{amsmath}
\usepackage{array,multirow}
\usepackage{amssymb}
\usepackage{graphicx}
\newcommand{\vect}[1]{{\boldsymbol{#1}}}

\def\cn{{\mathcal N}}

\def\la{\lambda}

\def\lb{\bar{\lambda}}

\def\ep{\epsilon}

\def\Or[#1]{{\text{O}}\left({#1}\right)}
\def\dotl[#1,#2]{\left\langle #1, #2 \right\rangle}
\def\dotlb[#1,#2]{[ #1, #2 ]}
\def\dotp[#1,#2]{(#1) \cdot (#2)}
\def\n4sym{{\cal N}=4 SYM}
\def\>{\rangle}
\def\<{\langle}
\def\ads[#1]{$\text{AdS}_{#1}$}
\def\lrdel{\overset{\leftrightarrow}{\nabla}}
\title{Recursion Relations for AdS/CFT Correlators}
\author{Suvrat Raju \\ Harish-Chandra Research Institute, \\ Jhunsi, Allahabad 211019.\\ E-mail: \email{suvrat@hri.res.in}}
\date{}
\abstract{
We expand on the results of \href{http://www.arxiv.org/abs/1011.0780}{arXiv:1011.0780} where we presented new recursion relations for  correlation functions of the stress tensor and conserved currents in conformal
field theories with an \ads[d+1] dual for $d \geq 4$. These recursion 
relations are derived by generalizing the Britto-Cachazo-Feng-Witten (BCFW)
relations to amplitudes in anti-de Sitter space (AdS) that
are dual to boundary correlators, and  are
usually computed perturbatively by Witten diagrams. Our results relate vacuum-correlation functions to integrated products of lower-point transition amplitudes, which correspond to correlators calculated between states dual to
 certain normalizable modes. 
We show that the set of ``polarization vectors'' for which amplitudes behave well under the BCFW extension is smaller than in flat-space. We describe how
transition amplitudes for more general external polarizations
can be constructed by combining answers obtained by different pairs of BCFW shifts. We then generalize these recursion relations to supersymmetric theories. In AdS, unlike flat-space, even maximal supersymmetry is insufficient to permit the computation of all correlators of operators in the same multiplet as a stress-tensor or conserved current. Finally, we work out some simple examples to verify our results.}
\preprint{HRI/ST/1103}

\begin{document}
\section{Introduction}
The Britto-Cachazo-Feng-Witten (BCFW) recursion relations \cite{Britto:2004ap,Britto:2005fq} have attracted interest because they provide a novel way of calculating S-matrix elements in gauge and gravity theories. These
relations reproduce the answers provided by standard Feynman-diagram
perturbation theory but are far more efficient. This provides an immediate
practical motivation for these investigations: the study of scattering
amplitudes that are of phenomenological relevance at particle accelerators. Indeed, using the BCFW relations and associated on-shell techniques
at one loop \cite{Britto:2004nc} -- \nocite{Bern:2005cq,Berger:2006ci,Forde:2007mi,ArkaniHamed:2008gz, Berger:2008sj,Ossola:2006us} \cite{Ellis:2007br}, Berger et al.  \cite{Berger:2010zx} were able to calculate the next-to-leading order correction 
for the production of W + 4-jets at the Large Hadron Collider (LHC)  ---  a calculation that had remained out of the reach of conventional Feynman diagram techniques. Earlier, two groups used similar techniques to understand the production of W + 3-jets at both the Tevatron and the LHC \cite{KeithEllis:2009bu,Melnikov:2009wh,Berger:2009zg}.

Another example, which is relevant to quantum gravity, comes from
scattering amplitudes of gravitons. If the Hilbert action is expanded 
in metric fluctuations, one obtains an infinite series of interaction
vertices of ever increasing complexity; for example, the four-point
vertex already has 2850 terms!  However, the interactions of 
gravitons are not {\em really} that complicated: the final answers
for scattering amplitudes are quite simple.  In fact, DeWitt
who first worked out scattering amplitudes for gravitons commented that ``the tediousness of the algebra involved \ldots combined with the fact that the final 
results are ridiculously simple, leads one to believe that there must
be an easier way.''\cite{DeWitt:1967uc} The BCFW recursion relations are this ``easier way''! They reduce the computation of all
 graviton scattering amplitudes down to the computation of the on-shell
three-point
function, which is completely determined by Lorentz invariance. 

This leads us to a second, more fundamental, reason for the interest
in the BCFW relations.  Since the 
BCFW recursion relations are ostensibly so different from Feynman diagrams 
but, yet, provide a shortcut to the final answer, we would like to find a 
formulation of quantum field theory (QFT) in which
these recursion relations, rather than Feynman diagrams, are the ``natural''
objects of study. A new such formulation of QFT, if it were
to be discovered, would certainly be important for our understanding
of basic physics.

The search for such a formulation has attracted much recent interest. An important
attempt was made by  Witten \cite{Witten:2003nn} who tried to formulate gauge theory as a string
theory in twistor space. Although this did not succeed entirely it led to the discovery
of the BCFW recursion relations. More-recent work that has aimed at uncovering 
the underlying physics in these relations includes a reformulation of 
the BCFW relations in twistor space \cite{ArkaniHamed:2009si,Mason:2009sa} and the 
attempt to find a Grassmannian integral \cite{ArkaniHamed:2009dn,Mason:2009qx} which would produce the leading singularities
of scattering amplitudes at all loops and the BCFW recursion relations at tree-level. For
other recent extensions to higher-loops we refer the reader to \cite{Gluza:2010ws,ArkaniHamed:2010kv}.

This programme of reformulating quantum field theory  is ambitious but has not yet succeeded. Nevertheless,
it is clear that the secret of the efficiency of the BCFW recursion 
relations, and on-shell techniques in general,  is that they make reference only to the physical degrees
of freedom in a theory. On the other hand, this comes at the price
of manifest locality. In a rough sense, this is analogous to the way
in which the path-integral formulation of QFT makes Lorentz-invariance
manifest but obscures unitarity; the Hamiltonian formulation, on the 
other hand, makes unitarity manifest but obscures Lorentz-invariance. 
The on-shell formulation that we are looking for would  make the simplicity of amplitudes 
and physical degrees of freedom manifest but perhaps obscure locality. 

All the studies that we have referred to above were carried out in flat-space. In fact,
until recently, it was believed that on-shell techniques 
apply only to quantum field theories in flat-space. This is because they rely heavily 
on the analytic properties of amplitudes, which change drastically in curved spacetime. 
In a recent letter \cite{Raju:2010by} we pointed out that the BCFW recursion relations could be generalized
to gauge and gravity theories in anti-de Sitter space (AdS). By
the AdS/CFT correspondence, this gives new and surprising recursion
relations for correlators in the boundary conformal field theory (CFT). 

The physical intuition underlying this surprising result is as follows. The BCFW recursion relations are predicated on the behaviour of Yang-Mills (YM) and gravity amplitudes when two of the external momenta are stretched off the infinity in a ``complex direction.'' Although this is not strictly 
a high energy limit, it is nonetheless true that the amplitude is dominated by interactions between a soft background and a highly boosted particle at a {\em single point} \cite{ArkaniHamed:2008yf}. In this limit, 
we do not expect this highly boosted particle to see the curvature of the neighboring spacetime region. Viewed from this perspective, the properties of an amplitude under
the BCFW extension should not change much as we go from flat-space to curved space, except
for one crucial difference: in flat-space, the location of this interaction-point does
not matter  whereas in curved space it does. So, in AdS we need to integrate over the different points where this interaction can occur. (This is similar to the intuition used in \cite{Polchinski:2001tt}.)  This process leads to the modified recursion relations that we present below. A higher-point correlator is broken down into the integral
of the product of two lower-point correlators. Just as in flat-space, we can continue this process till we are left only with three-point functions.

Once again, this is of interest for two reasons. First, the inordinate
complexity of gravity is exacerbated when we expand metric fluctuations
about AdS. As a result, even the smallest nontrivial correlators, like the 
four-point function of the stress-tensor in strongly coupled 
${\cn =4}$ Super-Yang-Mills theory (SYM) that is dual to the scattering
of four-gravitons in the bulk, have never been computed directly.\footnote{This correlator may, in principle, be extracted from simpler correlators using superconformal invariance \cite{Eden:2000bk,Drummond:2006by} but this has never been done explicitly either. On the other hand, as far as know, its direct computation using Witten diagrams has not even
been attempted.} Correlators of the stress-tensor are of particular interest 
because their leading behaviour is ``universal'' in any conformal field theory with a gravity
dual due to the fact that  tree-level graviton amplitudes are not sensitive to the matter-content 
of the theory. Since, as in flat-space, the repeated application of our recursion relations
allows us to reduce complicated amplitudes down to three-point functions, we expect
that our new recursion relations will greatly simplify the computation of these correlators. 
On the other hand, as we discuss in more detail in section \ref{secresults}, these relations are also of interest for formal reasons --- both for what they teach us about
quantum gravity in asymptotically anti-de Sitter spaces and for what they teach us about
conformal field theories with a gravity dual.

Finally, we should mention one interesting feature of our results. If we 
set out to compute 
a vacuum-correlator in the boundary theory, with all normalizable modes switched off in the bulk, the recursion relations
lead us to correlators computed in the presence of specific states; in the bulk, this corresponds to turning on some normalizable modes. We will
call these generalized correlators ``transition amplitudes.'' 

These transition amplitudes have a 
 nice physical interpretation in Lorentzian AdS. There has been much recent
discussion of the subtleties associated with Lorentzian AdS/ CFT; 
these subtleties are not too important here especially since we are 
are at zero temperature. However, the reader who is concerned about
this may instead prefer to work all the
time in the Euclidean picture and only analytically continue the results at the end.  In the Euclidean picture, the intermediate objects that we obtain in our recursion relations 
do not have any direct physical interpretation (except as the analytic continuation
of Lorentzian transition amplitudes) but they
are well defined formal quantities that one can compute in perturbation theory.

A brief overview of this paper is as follows. We start with a review of perturbation theory in AdS in section \ref{secreviewperturbation}. We then proceed
to define and discuss transition amplitudes in section \ref{sectransitionamp}.
 The central results in the paper are derived in section \ref{secbcfw} where we derive new recursion relations for transition amplitudes in Yang-Mills 
theory and gravity. 

A further extension of our recursion relations in section \ref{secsusytheories} allows us to compute transition
amplitudes in supersymmetric theories, including ${\cn=4}$ SYM and 
the theory on multiple M5 branes in the supergravity limit. Perturbative computations in supersymmetric theories are often tedious; the recursion relations 
that we present ameliorate this by using
a generalization of Nair's on-shell superspace \cite{Nair:1988bq}. We would 
suggest that the reader, who is interested just in the results of this paper, should
read \cite{Raju:2010by} first and then turn here for details.

\section{Review of Perturbation Theory  \label{secreviewperturbation}}
We will work
in Poincare coordinates where the metric is
\begin{equation}
\label{adsmetric}
d s^2 = g_{\mu \nu} d x^{\mu} d x^{\nu} = z^{-2} \left(d z^2 + \eta_{i j} d x^i d x^j \right).
\end{equation}
Note that we are using the mostly positive signature for the boundary metric. 
Poincare invariance in $d$-dimensions makes it convenient to Fourier
transform functions of $x^i$ and we will call the conjugate variables --- $k_i$ --- ``momenta.''These are really momenta in the dual conformal field theory. 
Note that we will use boldface for tensors and vectors in $d$ or $d+1$ dimensions, like $\vect{x}$ or $\vect{k}$, but not for their components. While considering momenta in a $n$-particle amplitude, we will often use the index $m$ to run
over the various particles from $1$ to $n$; these have momenta  from $\vect{k_1}$ to $\vect{k_n}$. At times, to lighten the notation, especially when 
we are dealing with ``polarization vectors'', $\vect{\ep_m}$, below we might raise the particle-number index i.e. write $\vect{\ep^m}$ instead; this does
not have any significance.

The indices, $i,j$ are reserved for the $d$-dimensional spacetime
coordinates. We will avoid raising and lowering these indices; for example
$x^i$ naturally has a raised index, while $k_i$ or a gauge field $A_i^{\rm a}$
naturally has lowered indices. However, when we need to take a dot product
of two $d$-dimensional vectors, we will use the flat-space metric. 
On the other hand, $\mu,\nu$ run over all $d+1$ dimensions. When we raise or lower one of these indices, we will use the full metric including the factors of $z$. Finally, the index $0$ will refer to the $z$-direction while the boundary coordinates run from $1 \ldots d.$

Perturbation theory in AdS is carried out through Witten diagrams. This 
requires two crucial ingredients: the bulk to boundary propagator, and the
bulk to bulk propagator. The bulk to boundary propagator is a certain kind
of solution to the equations of motion --- called a non-normalizable solution ---  with some special boundary conditions. We discuss these two physical
quantities for scalars, gauge fields and gravity below. These results
are well known but, at times, such as in the expression for the gauge and
gravity propagators, we were unable to find them in the literature in the 
exact-form that we required.  So, we have tried to be as detailed as
possible.
 
\subsection{Scalars}
We start by describing solutions to the wave equation and then go on to
describe propagators in AdS. 

\subsubsection{Solutions to the Wave Equation}
Consider a minimally coupled massless scalar. Its equation of motion is
\begin{equation}
\label{boxphizero}
\Box \phi = 0 \Rightarrow \partial_{\mu} g^{\mu \nu} \sqrt{-g} \partial_{\nu} \phi = 0.
\end{equation}
Poincare invariance in $d$-dimensions tells us that all solutions can be written as linear
combinations of  $\phi_{\vect{k}}(\vect{x},z) = e^{i \vect{k} \cdot \vect{x}} \phi(z)$
where $\phi_{\vect{k}}$ satisfies
\begin{equation}
\label{waveq}
\bigl(\partial_z z^{1 - d} \partial_z - z^{1-d} \vect{k}^2\bigr) \phi_{\vect{k}} = 0.
\end{equation}
Here, $\vect{k}^2 = \eta^{i j} k_i k_j$ is taken with the flat boundary metric. If $\vect{k}$ is timelike, which means $\vect{k}^2 < 0$, then there are two solutions to \eqref{waveq}
\begin{equation}
\label{timelikesolsphi}
\begin{split}
\text{normalizable:} \quad &\phi(z) = z^{\nu} \phi_0 J_{\nu}(|\vect{k}| z),\\
\text{non-normalizable:} \quad &\phi(z) = z^{\nu} \phi_0 Y_{\nu}(|\vect{k}| z),
\end{split}
\end{equation}
where $J_{\nu}$ and $Y_{\nu}$ are Bessel functions of the first and second kind
respectively. Moreover, $|\vect{k}| = \sqrt{|\vect{k}^2|}$, $\nu = {d \over 2}$ and\footnote{We hope this notation will not cause confusion with  the use of $\nu$ as a spacetime index.}  $\phi_0$ is some constant. Any linear combination of these solutions is
also a solution to the equation of motion. 

 On the other hand, if $\vect{k}$ is spacelike then the requirement that the solution be regular in the interior of AdS fixes the solution to be 
\begin{equation}
\label{spacelike}
\phi (z) = z^{\nu} \phi_0 K_{\nu}(|\vect{k}| z), \quad \text{for}~ \vect{k}^2 > 0,
\end{equation}
where $K$ is the modified Bessel function of the second kind. 

Note that the analytic continuation of \eqref{spacelike} to timelike momenta
gives a solution involving the Hankel function of the first kind\footnote{The factor of ${\pi \over 2} i^{\nu + 1}$ is customary in the definition of the modified Bessel functions.  We should
 also add here that we are not being very precise about the overall
normalization of these solutions since this is not
important for any part of the analysis in this paper}
\begin{equation}
\label{hankelsol}
\phi (z) = z^{\nu} \phi_0 {\pi \over 2} i^{\nu + 1} H_{\nu}^{(1)}(|\vect{k}| z) = z^{\nu} \phi_0 {\pi \over 2} i^{\nu + 1} \left(J_{\nu}(|\vect{k}| z) + i Y_{\nu}(|\vect{k}| z) \right), ~\text{for}~ \vect{k}^2 < 0.
\end{equation}
If we are calculating time-ordered correlation functions on the boundary,
then this is the correct bulk to boundary propagator \cite{Herzog:2002pc,Satoh:2002bc}. This is because, as we approach the Poincare horizon at $z = \infty$,
this bulk to boundary propagator ensures that positive energy waves are ingoing whereas 
negative energy waves are outgoing. 

In some contexts, we will find that the distinction between these solutions is unimportant. We will then write $\phi_{\vect{k}}(\vect{x}, z) = \phi_0 e^{i \vect{k} \cdot \vect{x}} E_{\nu} (\vect{k},  z)$ where $z^{-\nu} E_{\nu}(\vect{k}, z) $ is one of the Bessel functions, $K_{\nu}(\vect{k} z), J_{\nu}(\vect{k} z), Y_{\nu}(\vect{k} z)$, or a linear
combination of these functions. 

\subsubsection{Propagator}
We now turn to the bulk to bulk propagator. Here, we will just call this the propagator.  
The propagator is the Green's function that satisfies (in $d+1$ dimensions),
\begin{equation}
\label{bulkbulkdef}
\Box G(\vect{x}, z, \vect{x'},z') = i {\delta^d(\vect{x} - \vect{x'}) \delta(z-z') \over \sqrt{-g}}.
\end{equation}
Note that the left hand side is invariant under coordinate transformations
and the $\sqrt{-g}$ ensures that this is also true for the right hand side, because it
cancels the transformation of the $\delta$ function. (See Eqn. (3.49) of \cite{birrell1984quantum}.) Under a coordinate
transformation with Jacobian $J$, we have $\delta^d(x - x') \delta(z-z') \rightarrow \delta^d(x - x') \delta(z-z')/J$ and $\sqrt{-g} \rightarrow \sqrt{-g}/J$. 
With the metric \eqref{adsmetric} we have  $\sqrt{-g} = {1 \over z^{d+1}}$.

After we Fourier transform
from $\vect{x}$ to $\vect{k}$
\begin{equation}
\label{fourierGphi}
G_{\vect{k}}(z,z') = \int G(\vect{x}, z, \vect{x'}, z') e^{-i \vect{k} \cdot (\vect{x} - \vect{x'})} d^d x,
\end{equation}
the equation \eqref{bulkbulkdef} becomes 
\begin{equation}
\label{Geqn}
z^{d+1} {\partial \over \partial z} z^{1 - d} {\partial G_{\vect{k}} \over \partial z} - z^2 \vect{k}^2 G_{\vect{k}} = i \delta(z - z') z^{d + 1}.
\end{equation}
Using the identity, 
\begin{equation}
\label{hankel}
\int z J_{\nu}(p z) J_{\nu}(p' z) dz = {\delta (p - p') \over p},
\end{equation}
we see that the solution to \eqref{Geqn} is 
\begin{equation}
\label{Gscal}
G_{\vect{k}}(z,z') = \int {- i p \, d p \over 
\left(\vect{k}^2 + p^2 - i \epsilon \right)} z^{\nu} J_{\nu}(p z) J_{\nu} (p z') (z')^{\nu},
\end{equation}
and, as usual, 
\begin{equation}
\label{inversefourierG}
\begin{split}
G(\vect{x}, z, \vect{x'}, z') &= \int  {d^d \vect{k} \over (2 \pi)^d} G_{\vect{k}}(z,z') e^{i \vect{k} \cdot (\vect{x} - \vect{x'})} \\
&= \int {-i d^d \vect{k} \over (2 \pi)^d} {d p^2 \over 2}  {e^{i \vect{k} \cdot (\vect{x} - \vect{x'})} 
z^{\nu} J_{\nu}(p z) J_{\nu} (p z') (z')^{\nu} \over 
\left(\vect{k}^2 + p^2 - i \epsilon \right)},
\end{split}
\end{equation}

We draw the attention of the reader to one property of the momentum space propagator which will be important to us below. 
 When
the denominator of its integrand goes on shell i.e. when $p^2 = -\vect{k}^2$, the numerator
breaks up into a sum over a product of normalizable modes. This is the same
as what happens in flat-space, and is expected because the propagator is
just a two-point function. 

\subsection{Gauge Fields}
We now turn to vector fields in AdS. These are dual to conserved currents on the boundary. Both the solutions to the equations of motion and the propagator depend on the choice of gauge. Here, 
we will choose axial gauge so that for the gauge field $A^{\rm a}_{\mu}$ in the bulk: $A^{\rm a}_0 = 0.$ Note that we do not italicize the color-index ${\rm a}$.

The bulk action is 
\begin{equation}
\label{gaugeaction}
S = {-1 \over 4} \int \sqrt{-g} F_{\mu \nu}^{\rm a} F^{\mu \nu, \rm{a}} d^d \vect{x} d z.
\end{equation}
To go to ``axial gauge'', we add a gauge fixing term $\zeta \sum_{\rm a} (A^{\rm a}_0)^2$ and take
$\zeta \rightarrow -\infty$. This freezes $A^{\rm a}_0 = 0$. We also set the coupling constant, $g_{\text{YM}} = 0$ for now, although we will turn it on 
later to examine interactions in this theory.  With these choices, we have the 
gauge-fixed action
\begin{equation}
\label{axialactym}
\begin{split}
{S}_{\text{axial}} &= 
{1 \over 2} \int  \left[ A^{\rm a}_i \partial_{\mu} z^{3-d} \partial_{\rho} A^{\rm a}_{j} \eta^{\mu \rho} \eta^{i j} -   z^{3-d} A^{\rm a}_i \partial_k \partial_l A^{\rm a}_j \eta^{i k} \eta^{j l}\right] d^d \vect{x} d z + S_{\text{axial}}^{\cal B}.
\end{split}
\end{equation}
Note that we have used the fact that $A_{0}^{\rm a} = 0$ and also integrated by parts; $S_{\text{axial}}^{\cal B}$ is the resultant boundary term, which does not affect the equations of motion. 

From \eqref{axialactym}, we can read off both the solutions to the free-equations of motion and the propagator. 
Solutions to the equations of motion, 
\begin{equation}
\label{gaugesols}
A^{\rm a}_i(\vect{x},z) = \int A^{\rm a}_i(\vect{k},z) e^{+ i \vect{k} \cdot \vect{x}} d^d \vect{x},
\end{equation}
must satisfy
\begin{equation}
\label{transverseA}
\vect{k} \cdot \vect{A^{\rm a}}(\vect{k},z) = 0,
\end{equation}
in the gauge $A^{\rm a}_0(\vect{k},z) = 0$ and also 
\begin{equation}
\label{aieq}
\partial_0 z^{3-d} \partial_0 A^{\rm a}_i - k^2 z^{3-d} A^{\rm a}_i = 0.
\end{equation}
For timelike $\vect{k}$, this has the two solutions 
\begin{equation}
\label{gaugesoltime}
\begin{split}
\text{normalizable:} \quad A^{\rm a}_i(\vect{k}, z) &= \ep_{i}^{\rm a} z^{\nu_1} J_{\nu_1}(|\vect{k}| z), \\
\text{non-normalizable:} \quad A^{\rm a}_i(\vect{k}, z) &= \ep_{i}^{\rm a} z^{\nu_1} Y_{\nu_1}(|\vect{k}| z).
\end{split}
\end{equation}
Here $\nu_1 = \nu - 1$ and the polarization vector $\vect{\ep}$ must satisfy
\begin{equation}
\label{transversep}
\vect{k} \cdot \vect{\ep}^{\rm a} = 0.
\end{equation}
The timelike bulk to boundary propagator is a linear combination of these
solutions that has the correct boundary conditions at $z \rightarrow \infty$:
\begin{equation}
\label{gaugetimebulkbound}
A^{\rm a}_i(\vect{k}, z) = \ep_{i}^{\rm a} z^{\nu_1} H_{\nu_1}^{(1)}(|\vect{k}| z), ~\text{for}~ \vect{k}^2 < 0.
\end{equation}
On the other hand, for spacelike $\vect{k}$, we have the unique solution:
\begin{equation}
\label{gaugesolspace}
A^{\rm a}_i(\vect{k}, z) = \ep_{i}^{\rm a} z^{\nu_1} K_{\nu_1}(|\vect{k}| z), ~\text{for}~ \vect{k}^2 < 0.
\end{equation}
This is also the bulk to boundary propagator for spacelike momentum and
its analytic continuation to timelike momenta gives \eqref{gaugetimebulkbound}.
Inverting the quadratic operator in \eqref{axialactym} leads to the propagator:
\begin{equation}
\label{axialpropagator}
\begin{split}
&G^{\text{axial},{\rm a  b}}_{i j}(\vect{x}, z, \vect{x}',z') = \int {-i d^d \vect{k} d p^2 \over 2 (2 \pi)^d }  {e^{i \vect{k} \cdot (\vect{x} - \vect{x'})}}  \Bigl[{(z z')^{\nu_1} J_{\nu_1}(p z) J_{\nu_1} (p z') {\cal T}_{i j} \delta^{a b}\over 
\left(\vect{k}^2 + p^2 - i \epsilon \right)}\Bigr],
\end{split}
\end{equation}
where ${\cal T}_{i j} =  \eta_{i j} + {k_{i} k_{j} \over p^2}$
Once again we emphasize that at $\vect{k}^2 = -p^2$, ${\cal T}_{i j}$ just projects vectors onto
the space orthogonal to $\vect{k}$ and so the numerator of \eqref{axialpropagator} breaks up
into the sum over a product of normalizable modes.

\paragraph{Comparison with Liu-Tseytlin:}
We pause to compare our propagator to the one given by 
Liu and Tseytlin \cite{Liu:1998ty}.
Referring to Eqn. (4.6) of their paper (from version 4 on the arXiv), we see that, in the presence of a bulk-source $\vect{T}$, we can write the quadratic action as 
\begin{equation}
\label{ltact}
\begin{split}
2 S_{LT}&=\int {
d z d z' 
{\cal D}(\vect{k},\vect{x},\vect{x'})
 d p^2 \over 2 z^{d+1} (z')^{d+1}}  T^i(z) \left[{\eta_{i j} - {k_i k_j \over \vect{k}^2} \over \vect{k}^2 + p^2}\right] T^j(z') z^{\nu_1} 
J_{\nu_1}(|\vect{k}|z) (z')^{\nu_1} J_{\nu_1}(|\vect{k}| z') 
 \\
&+ \int T^0(\vect{x},z) T^0(\vect{x'},z) 
{1 \over \vect{k}^2} {{\cal D}(\vect{k},\vect{x},\vect{x'}) d z \over z^{d+5}}.
\end{split}
\end{equation}
Note that (a) we have adopted the notation 
\begin{equation}
{\cal D}(\vect{k},\vect{x}, \vect{x'}) \equiv d^d \vect{x} d^d \vect{x'}  {d^d \vect{k} \over (2 \pi)^d} e^{i \vect{k} \cdot (\vect{x} - \vect{x'})},
\end{equation}
(b) we have raised indices for the currents, which we have denoted by $\vect{T}$ to avoid
confusion with Bessel functions; this leads to slightly different
factors of z. Moreover, we have only displayed the $z$ dependence in the 
source which may also depend on $\vect{x}$. (c) We have a minus sign in the second
term by virtue of having $\vect{k}^2$ rather than $\partial^2$  and (d) $\nu_1 = {d - 2 \over 2}$.  
On the other hand, we also have
\begin{equation}
\label{axialact}
\begin{split}
&2 S_{\rm axial} =\int {d z d z' {\cal D}(\vect{k},\vect{x}, \vect{x'})  d p^2 \over 2  z^{d+1} (z')^{d+1}} 
T^i(z) \left[{\eta_{i j} + {k_i k_j \over p^2} \over \vect{k}^2 + p^2} \right] T^j(z') z^{\nu_1} J_{\nu_1}(|\vect{k}| z) (z')^{\nu_1} J_{\nu_1}(|\vect{k}|z'),
\end{split}
\end{equation}
so that
\begin{equation}
\label{diffbetact}
\begin{split}
&2 S_{LT} - 2 S_{\rm axial} \\
&= \int {d z d z' {\cal D}(\vect{k},\vect{x}, \vect{x'})  d p^2 \over 2 z^{d+1} (z')^{d+1}} T^i(z) \left[{{-k_i k_j \over k^2} - {k_i k_j \over p^2} \over k^2 + p^2}\right] T^j(z') z^{\nu_1} J_{\nu_1}(|\vect{k}| z) (z')^{\nu_1} J_{\nu_1}(|\vect{k}|z') \\
&+ \int T^0(z) T^0(z) 
{1 \over \vect{k}^2} {{\cal D}(\vect{k},\vect{x}, \vect{x'}) d z \over z^{d+5}} \\
&=  \int {d z d z' {\cal D}(\vect{k},\vect{x}, \vect{x'} ) d p^2 
\over 2 z^{d+1} (z')^{d+1} } T^i(z) \left[{-k_i k_j \over \vect{k}^2 p^2} \right] T^j(z') z^{\nu_1} J_{\nu_1}(|\vect{k}| z) (z')^{\nu_1} J_{\nu_1}(\vect{x'},z') 
\\
&+ \int T^0(z) T^0(z) 
{
1 \over \vect{k}^2} {{\cal D}(\vect{k},\vect{x}, \vect{x'}) d z \over z^{d+5}}.
\end{split}
\end{equation}
We now use the identities
\begin{equation}
\label{besselintegral}
\begin{split}
&\int_0^{\infty} J_{\nu_1}(p z) J_{\nu_1}(p z') {d p \over p} = {\theta(z - z') \over 2 \nu_1} \left({z' \over z}\right)^{\nu_1} +  {\theta(z' - z)\over 2 \nu_1} \left({z \over z'}\right)^{\nu}, \\
&\partial_i T^i = -{1 \over \sqrt{-g}} \partial_0 \sqrt{-g} T^0,
\end{split}
\end{equation}
where the second line comes from current conservation and is useful
because we can replace  $k_i k_j \rightarrow \partial_i \partial_j'$. Note that one derivative pulls down an $i$ and the other pulls
down a $(-i)$, so there is no overall minus sign. Substituting
this into the equation above, and integrating by parts, we see that
\begin{equation}
\label{diffact2}
\begin{split}
&2 (S_{LT} - S_{\rm axial})\\
 &= -\int {d z d z' {\cal D}(\vect{k},\vect{x}, \vect{x'})  \over z^{d+1} (z')^{d+1} } 
{1 \over  k^2} T^0(z)T^0(z')  {\cal P}_{\theta, z, z'}
  + \int T^0(z) T^0(z) {{\cal D}(\vect{k},\vect{x}, \vect{x'}) \over k^2} {d z \over z^{d+5}}.
\end{split}
\end{equation}
where we have defined
\begin{equation}
{\cal P}_{\theta, z, z'} \equiv {\partial^2 \over \partial z \partial z'}\left({z^{\nu_1} (z')^{\nu_1} \over 2 \nu_1} \left[\theta(z - z') \left({z' \over z}\right)^{\nu_1} +  \theta(z' - z) \left({z \over z'}\right)^{\nu_1} \right]\right).
\end{equation}
Carefully working out the derivative, we find
\begin{equation}
\label{derivativetheta}
\begin{split}
{\cal P}_{\theta, z, z'} &= {\partial \over \partial z'} \left[{\partial \over \partial z} \left((z')^{2 \nu_1} \theta(z - z') + z^{2 \nu_1} \theta(z' - z) \right)\right] \\ &= {\partial \over \partial z'} \left[(z')^{2 \nu_1} \delta(z - z') - z^{2 \nu_1} \delta(z' - z) + {2 \nu_1} z^{2 \nu_1 - 1} \theta(z' - z) \right]\\
&= {\partial \over \partial z'} {2 \nu_1} z^{2 \nu_1 - 1} \theta(z' - z) = 2 \nu_1 z^{2 \nu_1 - 1} \delta(z' - z).
\end{split}
\end{equation}
Substituting this into \eqref{diffact2}, we see that everything cancels out
miraculously and we get 
\begin{equation}
\label{bothequal}
S_{LT} - S_{\rm axial} = 0.
\end{equation}
So, our propagator is the same as the propagator given by Liu and Tseytlin
although the two are written in slightly different forms.

\subsection{Gravity}
Now, we turn to gravitons propagating in AdS. We expand 
gravity fluctuations about a background metric ${\mathcal{G}}_{\mu \nu} = g_{\mu \nu} + h_{\mu \nu}$, where $g_{\mu \nu}$ is the background and $h_{\mu \nu}$ contains the 
fluctuations. We can take the quadratic
gravity action from \cite{Christensen:1979iy}. As in the subsection above,
boundary terms affect neither the equations of motion nor the propagator,
so we will neglect them here. 
The quadratic action is given by
\begin{equation}
\label{gravityquadratic}
S = {-1 \over 64 \pi G} \int {d^d \vect{x} d z \over z^{d+1}} \left(\tilde{h}^{\mu \nu} \Box h_{\mu \nu} + 2 \tilde{h}^{\mu \nu} R_{\mu \rho \nu \sigma} h^{\rho \sigma}  + 2 \nabla^{\rho} \tilde{h}_{\rho \mu} \nabla^{\sigma} \tilde{h}_{\sigma}^{\mu} \right),
\end{equation}
where $\tilde{h}^{\mu \nu} = h^{\mu \nu} - {1 \over 2} g^{\mu \nu} h^{\alpha \beta} g_{\alpha \beta}$, 
and all covariant 
derivatives are with respect to the background metric.

We want to analyze this action, when the background metric is the AdS metric, in axial gauge $h_{\mu 0} = 0.$ 
Before we parse this
action, let us write down a few simple identities. The connection
coefficients are given by
\begin{equation}
\label{connectionval}
\Gamma^{\rho}_{\alpha \beta} = {1 \over 2} g^{\rho \delta} \left(\partial_{\alpha} g_{\beta \delta} + \partial_{\beta} g_{\alpha \delta} - \partial_\delta g_{\alpha \beta} \right) =  {1 \over z} \left(\delta^{\rho}_{0} \eta_{\alpha \beta} - \delta^{0}_{\alpha} \delta^{\rho}_{\beta} - \delta^{0}_{\beta} \delta^{\rho}_{\alpha} \right),
\end{equation}
Using these coefficients, we see that 
\begin{equation}
\label{covardertensor}
\begin{split}
\nabla_{\rho} h_{\mu \nu} &= \partial_{\rho} h_{\mu \nu} - \Gamma^{\alpha}_{\rho \mu} h_{\alpha \nu} - \Gamma^{\alpha}_{\rho \nu} h_{\mu \alpha}\\
&= \partial_{\rho} h_{\mu \nu} + {1 \over z} \left(2 \delta^{0}_{\rho} h_{\mu \nu} + \delta^{0}_{\mu} h_{\rho \nu} + \delta^0_{\nu} h_{\mu \rho} \right) \\
&= {1 \over z^2} \left(\partial_{\rho} (z^2 h_{\mu \nu}) + z \delta^{0}_{\mu} h_{\rho \nu} + z \delta^0_{\nu} h_{\mu \rho} \right).
\end{split}
\end{equation}
In particular, in the action, we have a term of the kind
\begin{equation}
\label{simpletwoderiv}
\begin{split}
&\sqrt{-g}g^{\gamma \rho} g^{\mu \mu_1} g^{\nu \nu_1} \nabla_{\gamma} h_{\mu_1 \nu_1} \nabla_{\rho} h_{\mu \nu} \\
&= z^{1-d} \eta^{\gamma \rho} \eta^{\mu \mu_1} \eta^{\nu \nu_1} \left[\partial_{\gamma}(z^2 h_{\mu_1 \nu_1}) + z \delta^0_{\mu_1}  h_{\gamma \nu_1} + z \delta^{0}_{\nu_1} h_{\mu_1 \gamma} \right] \left[\partial_{\rho} (z^2 h_{\mu \nu}) + \delta^0_{\mu} z h_{\rho \nu} + z \delta^0_{\nu} h_{\mu \rho} \right].
\end{split}
\end{equation}
Note that when we expand this product out, the cross-terms all contract to zero with our choice
of gauge.  However, we are left with
\begin{equation}
\label{simpletwoderiv1}
\sqrt{-g}g^{\gamma \rho} g^{\mu \mu_1} g^{\nu \nu_1} \nabla_{\gamma} h_{\mu_1 \nu_1} \nabla_{\rho} h_{\mu \nu} = z^{1-d}   \eta^{\mu \mu_1} \eta^{\nu \nu_1} \left[\eta^{\gamma \rho} \partial_{\gamma}(z^2 h_{\mu_1 \nu_1}) \partial_{\rho} (z^2 h_{\mu \nu}) + 2 z^2 h_{\mu \nu} h_{\mu_1 \nu_1} \right].   
\end{equation}
If we integrate this by parts, we find 
\begin{equation}
\label{simpletwoderiv2}
\begin{split}
&\int_{\vect{x},z} \sqrt{-g}g^{\gamma \rho} g^{\mu \mu_1} g^{\nu \nu_1} \nabla_{\gamma} h_{\mu_1 \nu_1} \nabla_{\rho} h_{\mu \nu} \\
\sim   &\int_{\vect{x},z} \eta^{\mu \mu_1} \eta^{\nu \nu_1} \left[ -z^2 \eta^{\gamma \rho} h_{\mu \nu} \partial_{\rho} z^{1-d} \partial_{\gamma}(z^2 h_{\mu_1 \nu_1}) + 2 z^{3-d}  h_{\mu \nu} h_{\mu_1 \nu_1} \right],
\end{split}
\end{equation}
where $\sim$ indicates that the equality holds up to boundary terms that
are unimportant for our purpose and we have adopted the notation $\int_{\vect{x},z} \equiv \int d^d \vect{x} \, d z.$
Note that in contrast, for a scalar, the Laplacian is just
\begin{equation}
\label{scallaplacian}
\sqrt{-g} \Box h = \eta^{\rho \gamma} \partial_{\rho} z^{1-d} \partial_{\gamma} h.
\end{equation}
The action in \eqref{gravityquadratic} also has a term that reads
\[
\label{crossterm}
{-1 \over 2} \nabla_{\rho} \tilde{h}_{\mu \nu} \nabla_{\sigma} \tilde{h}_{\mu_1 \nu_1} g^{\mu \rho} g^{\nu \nu_1} g^{\sigma \mu_1},
\]
with $\tilde{h}_{\mu \nu} = h_{\mu \nu} - {1 \over 2} h g_{\mu \nu}$. When we expand this term out, we get three-types of terms: tensor-tensor, scalar-scalar and tensor-scalar. Let us look at these terms in a little 
more detail. The tensor-tensor term is 
\begin{equation}
\label{tentenextra}
\begin{split}
&{-\sqrt{-g} \over 2} \nabla_{\rho} {h}_{\mu \nu} \nabla_{\sigma} {h}_{\mu_1 \nu_1} g^{\mu \rho} g^{\nu \nu_1} g^{\sigma \mu_1} \\
&= 
{- z^{1-d} \over 2 } \left[ \partial_{\rho}(z^2 h_{\mu \nu}) + z \delta^0_{\mu} h_{\rho \nu} + z \delta^0_{\nu} h_{\mu \rho} \right]  \left[ \partial_{\sigma}(z^2 h_{\mu_1 \nu_1}) + z \delta^0_{\mu_1} h_{\sigma \nu_1} + z \delta^0_{\nu_1} h_{\mu_1 \sigma} \right]  \eta^{\mu \rho} \eta^{\nu \nu_1} \eta^{\sigma \mu_1} \\
&= {- z^{1-d} \over 2 } \left[  \eta^{\nu \nu_1}(i k_{\rho})(z^2 h_{\mu \nu}) (i k_{\sigma})(z^2 h_{\mu_1 \nu_1})  +  z^2 h_{\mu \rho} h_{\mu_1 \sigma} \right]   \eta^{\mu \rho} \eta^{\sigma \mu_1}. \\
\end{split}
\end{equation}

There are two tensor-scalar terms, so we have added a factor of two below. 
However, there is another factor of ${1 \over 2}$ from the coefficient of the
scalar in the definition of $\tilde{h}$.
We also have an overall positive sign because the minus sign in the coefficient
and the relative minus sign between the tensor and scalar cancel.
The tensor-scalar term is now
\begin{equation}
\label{tenscal}
\begin{split}
&{\sqrt{-g} \over 2} \nabla_{\rho} {h}_{\mu \nu} \nabla_{\sigma} {h} g_{\mu_1 \nu_1} g^{\mu \rho} g^{\nu \nu_1} g^{\sigma \mu_1} \\
&= {1 \over 2} z^{1-d} \left[ \partial_{\rho}(z^2 h_{\mu \nu}) + z \delta^0_{\mu} h_{\rho \nu} + z \delta^0_{\nu} h_{\mu \rho} \right] \partial_{\sigma} (z^2 h_{\mu_2 \nu_2} \eta^{\mu_2 \nu_2}) \eta^{\nu \sigma} \eta^{\mu \rho} \\
&= {1 \over 2} z^{1-d} \left[z^4 (i k_{\rho}) (i k_{\sigma}) h_{\mu \nu} \eta^{\rho \mu} \eta^{\nu \sigma} h_{\mu_2 \nu_2} \eta^{\mu_2 \nu_2} + h_{\mu \rho} \eta^{\mu \rho} z \partial_{z} (z^2 h_{\mu_2 \nu_2} \eta^{\mu_2 \nu_2}) \right].
\end{split}
\end{equation}
Writing $h = z^2 \eta^{\mu \rho} h_{\mu \rho}$, the second term in the bracket
above is 
\begin{equation}
\label{scscextra}
{1 \over 2} z^{-d} h \partial_z h = {1 \over 4} z^{-d} \partial_z h^2 \sim h^2 {{d} \over 4} z^{-{(d+1)}},
\end{equation}
where we have integrated by parts to get the last term.
Putting \eqref{scscextra} and \eqref{tentenextra} together, we find that
\begin{equation}
\label{sumtetenplusscs}
\begin{split}
&\int_{\vect{x},z} {-\sqrt{-g} \over 2} \nabla_{\rho} \tilde{h}_{\mu \nu} \nabla_{\sigma} \tilde{h}_{\mu_1 \nu_1} g^{\mu \rho} g^{\nu \nu_1} g^{\sigma \mu_1} \\
&= \int_{\vect{x},z} \Bigl[{z^{5-{d}} \over 2 } \eta^{\nu \nu_1} \eta^{\mu \rho} \eta^{\sigma \mu_1} k_{\rho} h_{\mu \nu}  k_{\sigma} h_{\mu_1 \nu_1} - {z^{5-{d}} \over 2} k_{\rho} k_{\sigma} h_{\mu \nu} h_{\mu_2 \nu_2}  \eta^{\rho \mu} \eta^{\nu \sigma}  \eta^{\mu_2 \nu_2} \\
&\hphantom{= \int_{\vect{x},z} \Bigl[}
+ {1 \over 8 z^{d+1}} h \Box h  + {({d} - 2) \over 4 z^{d+1}} h^2 \Bigr]. 
\end{split}
\end{equation}

We now add the simple contribution from the Riemann tensor to \eqref{sumtetenplusscs} and \eqref{simpletwoderiv2}. This allows us 
to derive solutions to the equations of motion and, by inverting the quadratic part of 
this action, we also obtain the propagator in axial gauge. 
In this gauge, the solutions to the equations of motion
are given by transverse traceless tensors in $d$-dimensions
\begin{equation}
\label{normalsolgrav}
h_{i j} =  \ep_{i j} z^{-2} E_{\nu}(\vect{k}, z) e^{i \vect{k} \cdot \vect{x}}; \quad h_{0\mu} = 0, \quad k_i \ep^{i j} = 0, \quad \ep^{i}_{i} = 0.
\end{equation}
The propagator is
\begin{equation}
\label{gravitypropagator}
G^{{\text{grav}}}_{i j, k l} =
\int \left[{
e^{i \vect{k} \cdot (\vect{x} - \vect{x'})} 
z^{\nu-2} J_{\nu}(p z) J_{\nu} (p z') (z')^{\nu - 2} \over  
\left(\vect{k}^2 + p^2 - i \epsilon\right)} \right.   {1 \over 2} \left.\left({\cal T}_{i k} {\cal T}_{j l} + {\cal T}_{i l} {\cal T}_{j k} - 
{2 {\cal T}_{i j} {\cal T}_{k l}\over d-1} \right)\right] {-i d^d \vect{k} d p^2 \over 2 (2 \pi)^d},  
\end{equation}
where ${\cal T}_{i j} = \eta_{i j} + k_{i} k_{j}/p^2$. A comparison similar
to the one done above for the gauge field shows that this agrees with 
the propagator given in \cite{Liu:1998ty} although it is written in 
a different form.

\section{Transition Amplitudes \label{sectransitionamp}}
The AdS/CFT prescription relates a field $\phi$ in the bulk
to an operator $O$ on the boundary and states that 
\begin{equation}
\label{adscft}
\bigl. \int_{\text{AdS}} e^{-S} \bigr|_{\phi(z,\vect{x}) \underset{z \rightarrow 0}{\longrightarrow} \phi_0(\vect{x})} = \langle e^{\int \phi_0(\vect{x}) O(\vect{x}) d^d \vect{x}} \rangle_{\text{CFT}}.
\end{equation}
By differentiating the right and left hand sides with respect to $\phi_0$, 
we get CFT correlators of $O$ on the right hand side and Witten diagrams
on the left hand side, in the limit where the string theory in AdS can be 
treated perturbatively. For a review of Witten diagrams, see \cite{D'Hoker:2002aw}.

The usual correlators we get in this manner are vacuum correlators, which is 
what the Euclidean path integral naturally calculates. In this paper, 
it will be physically more illuminating to consider the Lorentzian analogue
of \eqref{adscft} (which must be carefully defined \cite{Herzog:2002pc})
and consider correlators evaluated {\em between} states. More precisely, 
consider CFT operators $O(\vect{k_{3 1}}), \ldots O(\vect{k_{3 n_3}})$ and states $s, s'$ that are dual, respectively, to linear combinations 
of normalizable modes with momenta $\vect{k_{1 1}}, \ldots \vect{k_{1 n_1}}$ and $\vect{k_{2 1}}, \ldots \vect{k_{2 n_2}}$ in
the bulk. 
An important object in our study will be the 
transition amplitude
\begin{equation}
\label{transdef}
T(\vect{k_{l m}}) (2 \pi)^{d} \delta^{d}(\sum_{l m} \vect{k_{l m}}) = \langle s | O(\vect{k_{3 1}}) \ldots O(\vect{k_{3 n_3}}) | s' \rangle.
\end{equation}
We have an overall momentum-conserving delta-function because of translational invariance
on the boundary. We have explicitly extracted this in the definition of $T$ above. Physically, we may 
think of $|s'\rangle, \langle s |$ as specifying data along the past and future horizons of the Poincare patch; we are then asking for the probability that
the operators $O(\vect{k_{3 m}})$ will induce a transition between these states. Since   $|s'\rangle, \langle s |$ are dual to classical solutions in the bulk, these are coherent states.

Transition amplitudes are not usually considered in the literature although
they were discussed briefly in \cite{Balasubramanian:1999ri,Balasubramanian:1998de,Balasubramanian:1998sn}. Nevertheless, they are very natural objects
to compute in perturbation theory. The perturbative prescription for 
computing them is as follows.
We
draw bulk-bulk diagrams as usual. Then we contract the legs with momenta in the set $\vect{k_{3 m}}$ with
bulk to boundary propagators (non-normalizable modes), and the other legs, which carry  momenta in the set $\vect{k_{1 m}}$ or $\vect{k_{2 m}}$, with normalizable modes. So, a transition amplitude is merely obtained by replacing some of the bulk-to-boundary legs of a Witten diagram with normalizable modes. A vacuum-correlator is, of course, just a special case of a transition amplitude where all
normalizable modes are switched off.

It was pointed out in \cite{Balasubramanian:1999ri}
that a transition amplitude in the Poincare patch
may be thought of as a correlation function in global AdS. 
\begin{figure}
\label{poincaretoglobal}
\begin{center}
\includegraphics[height=0.4\textheight]{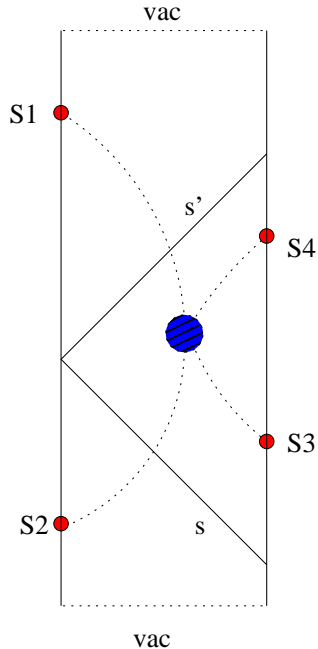}
\caption{Poincare Transition Amplitudes as Vacuum Correlators in global AdS}
\end{center}
\end{figure}
To see this, consider computing a Witten diagram in global AdS, as shown in Fig. \ref{poincaretoglobal}, with sources S1,S2,S3,S4, and the initial and final
boundary conditions set to the vacuum. This is evidently a four-point 
vacuum-correlator in the boundary theory.
From the point of view of the Poincare patch, however, the sources S1 and S2 
are invisible. Their effect is to create some boundary conditions on the 
past and the future horizons. This is precisely a transition amplitude.

The problem with this analogy, however, is that if we want to correctly
compute the correlation function in global AdS, we need to integrate over all points where the interaction takes place {\em including} points that are outside the Poincare patch. So, the analogy is strictly correct only if we use
sources that are ``aimed'' to allow for interactions within the Poincare patch. For this reason --- although this point of view is conceptually important --- we will not follow this analogy further since it is
simpler to deal directly with the perturbative prescription for transition
amplitudes.  

Below, we will consider transition amplitudes, $T(\vect{k_m}, \vect{\ep^{\rm a_m}})$, that depend on a set of discrete momenta, and also on polarization vectors for gauge-bosons and gravitons. The reader should note that some of the $\vect{k_m},\vect{\ep^{\rm a_m}}$ may correspond to normalizable modes, and others to non-normalizable modes. Since we wish to treat these cases symmetrically,
our notation will leave this implicit, although the context should suffice
to prevent any confusion.

It is clear from this discussion that  our transition amplitudes have a nice physical interpretation in 
Lorentzian AdS. On the other hand we can also 
consider the formulae below in Euclidean space --- where transition amplitudes continue to 
be well defined formal objects in perturbation theory ---  and then analytically continue the results
to Lorentzian space. 

\subsection{Ward Identities}
Transition amplitudes in Yang-Mills theory or gravity in AdS obey Ward identities,
just like flat-space S-matrix elements. To see the form
 that these identities take, let us note that the structure  of perturbation theory tells us that transition amplitudes
are produced by the action of a 
multilinear operator on a set of (normalizable or non-normalizable) solutions to the equations of motion.
For example, in Yang-Mills theory, with $A^{\rm a_m}_{\mu_m}(\vect{x},z)$ drawn from \eqref{gaugesoltime} or \eqref{gaugesolspace}
\begin{equation}
\label{multilinearop}
T = G(A^{\rm a_1}_{\mu_1}(\vect{x},z), \ldots A^{\rm a_n}_{\mu_n}(\vect{x},z)).
\end{equation}

In flat-space we usually do not have to think of multilinear operators acting on
the equations of motion --- the S-matrix element is simply some tensor, which comes from a sum of amputated Feynman diagrams,  dotted with the polarization vectors. Here, since we have not Fourier transformed with respect to the $z$-coordinate, the ``amputated'' Green's function could contain derivatives in $z$ that can 
act on the $z$-dependent pieces of the solutions to the equations of motion. This is why we need to consider linear operators that are more general than tensors here.

In Yang-Mills theory, these operators obey Ward identities:
\begin{equation}
\label{wardidentity}
G(\nabla_{\mu_1} \phi^{\rm a_1}(\vect{x},z), A^{\rm a_2}_{\mu_2}(\vect{x},z), \ldots A^{\rm a_n}_{\mu_n}(\vect{x},z)) = 0,
\end{equation}
for any  $\phi^{\rm a_1}(\vect{x},z).$ 
For gravity, these Ward identities can be written
\begin{equation}
\label{wardidentitygravity}
G(\nabla_{(\mu_1} t_{\nu_1)}(\vect{x},z), h_{\mu_2 \nu_2}(\vect{x},z), \ldots h_{\mu_n \nu_n}(\vect{x},z)) = 0,
\end{equation}
for any  vector field $t_{\nu_1}(\vect{x},z).$ 

\section{BCFW Recursion \label{secbcfw}}
Consider a $n$-point transition amplitude with momenta $\vect{k_1}, \ldots \vect{k_n}.$ We choose a $d$-dimensional null-vector $\vect{q}$, which has the property that
\begin{equation}
\vect{q} \cdot \vect{k_1} = \vect{q} \cdot \vect{k_n} = \vect{q}^2 = 0.
\end{equation}
$\vect{q}$ is not unique but, in general, some of its components will be complex.  With some choice of $\vect{q}$, we now consider a one-parameter momentum-conserving deformation of 
the transition amplitude, which we implement via
\begin{equation}
\label{bcfwextension}
\vect{k_1} \rightarrow \vect{k_1} + \vect{q} w, \quad  \vect{k_n} \rightarrow \vect{k_n} - \vect{q} w,
\end{equation}
where $w$ is a complex parameter.
We will examine in turn, what happens to transition amplitudes involving scalars, gauge bosons and gravitons under this extension.

\subsection{Scalars}
We start with a massless $\phi^3$ theory because its perturbation theory is simple and illustrative. Consider a four-point vacuum
correlator in this theory. There are three terms that contribute to this.\footnote{The $z$-integrals in \eqref{typical} need to be regulated, but this does not affect our analysis. We discuss this briefly at the end of section \ref{secexamples}.}
\begin{equation}
\label{typical}
\begin{split}
&T(\vect{k_1}, \vect{k_2}, \vect{k_3}, \vect{k_4}) =
\int  {i d z_1 d z_2 d p^2 \over 2 (z_1 z_2)^{d+1}} \\ \times &\Bigl[{z_1^{\nu} H_{\nu}^{(1)}(|\vect{k_1}| z_1) z_1^{\nu} H_{\nu}^{(1)}(|\vect{k_2}| z_1)  z_1^{\nu} J_{\nu}(p z_1)   z_2^{\nu} J_{\nu}(p z_2) z_2^{\nu} H_{\nu}^{(1)}(|\vect{k_3}| z_2) z_2^{\nu} H_{\nu}^{(1)}(|\vect{k_4}| z_2)  \over (\vect{k_1} + \vect{k_2})^2 + p^2}  \\
&+ {z_1^{\nu} H_{\nu}^{(1)}(|\vect{k_1}| z_1) z_1^{\nu} H_{\nu}^{(1)}(|\vect{k_3}| z_1)  z_1^{\nu} J_{\nu}(p z_1)   z_2^{\nu} J_{\nu}(p z_2) z_2^{\nu} H_{\nu}^{(1)}(|\vect{k_2}| z_2) z_2^{\nu} H_{\nu}^{(1)}(|\vect{k_4}| z_2)  \over (\vect{k_1} + \vect{k_3})^2 + p^2} \\
&+ {z_1^{\nu} H_{\nu}^{(1)}(|\vect{k_1}| z_1) z_1^{\nu} H_{\nu}^{(1)}(|\vect{k_4}| z_1)  z_1^{\nu} J_{\nu}(p z_1)   z_2^{\nu} J_{\nu}(p z_2) z_2^{\nu} H_{\nu}^{(1)}(|\vect{k_3}| z_2) z_2^{\nu} H_{\nu}^{(1)}(|\vect{k_2}| z_2)  \over (\vect{k_1} + \vect{k_4})^2 + p^2} \Bigr],
\end{split}
\end{equation} 
where $\nu = {d \over 2}$ as usual. The first point to note is that if we 
extend $\vect{k_1}$ and $\vect{k_4}$ using \eqref{bcfwextension}, the analytic properties of the {\em integral} \eqref{typical} in the $w$ plane are quite complicated.  This might, at first sight, seem like an obstruction to the use of the BCFW
recursion relations. However, the key point is that the {\em integrand} of \eqref{typical} is a rational function of $w$. Note that this is crucially dependent on the fact that the BCFW extension does not change the norm of $\vect{k_1}$ or $\vect{k_4}$, and so does not affect the Bessel function. A rational function can be reconstructed from a knowledge of its behaviour at infinity, the location of its poles and its residues there. 

The first two terms (which we call the s- and t-channel terms) inside the square brackets of \eqref{typical} have poles at finite $w$:
\begin{equation}
\begin{split}
w_s(p) &= -{(p^2 + (\vect{k_1} + \vect{k_2})^2) \over  (2 \vect{q} \cdot \vect{k_2})}, \\
w_t(p) &=  -{(p^2 + (\vect{k_1} + \vect{k_3})^2) \over  (2 \vect{q} \cdot \vect{k_3})}.
\end{split}
\end{equation}
We emphasize that the position of these poles depends on the value of $p$. The residues at these poles are 
\begin{equation}
\label{residue}
\begin{split}
R_s = {-i \over  4 \vect{q} \cdot \vect{k_2}} &\times \left[-i z_1^{3 \nu - d - 1} H_{\nu}^{(1)}(|\vect{k_1}| z_1) H_{\nu}^{(1)}(|\vect{k_2}| z_1)  J_{\nu}(p z_1)\right] \\ &\times \left[ -i z_2^{3 \nu - d - 1} J_{\nu}(p z_2)  H_{\nu}^{(1)}(|\vect{k_3}| z_2) H_{\nu}^{(1)}(|\vect{k_4}| z_2) \right] ,\\ 
R_t = {-i \over  4 \vect{q} \cdot \vect{k_3}} &\times \left[-i z_1^{3 \nu - d - 1} H_{\nu}^{(1)}(|\vect{k_1}| z_1) H_{\nu}^{(1)}(|\vect{k_3}| z_1) J_{\nu}(p z_1) \right] \\ &\times \left[ -i z_2^{3 \nu - d - 1} J_{\nu}(p z_2) H_{\nu}^{(1)}(|\vect{k_2}| z_2) H_{\nu}^{(1)}(|\vect{k_4}| z_2) \right] .
\end{split}
\end{equation}
However, these residues have another nice feature. The boundary momentum that
runs through the s-channel propagator is $\vect{k'_{s}} = -\vect{k_1} - \vect{k_2} - \vect{q} w_s$.  Note that, by construction, at $w = w_s$, we have $p^2 = |\vect{k'_s}|^2$. Therefore, each bracketed terms is the integrand for a 3-point function! 

There is
also a pole at $w = \infty$ in \eqref{typical} because the integrand of the $u$-channel term, which involves a 
contact interaction between $\vect{k_1}$ and $\vect{k_4}$, goes to a constant at large $w$. The residue at this pole cannot be written as the product of the
integrands of three-point amplitudes and it must be computed explicitly. If we
denote the value of this Witten diagram by ${\cal B}$ (this is the ``boundary term'' from $w = \infty$), then
we see that the following relation holds
\begin{equation}
\begin{split}
&T(\vect{k_1}, \vect{k_2}, \vect{k_3}, \vect{k_4}) = {\cal B} + \int \left[{-i {\cal T}_s^2 \over 2(p^2 + (\vect{k_1} + \vect{k_2})^2)} + {-i {\cal T}_t^2 \over 2(p^2 + (\vect{k_1} + \vect{k_3})^2)}\right]  d p^2, \\
&{\cal T}_s^2 \equiv  {T(\vect{k_1 + \vect{q} w_s(p)},k_2,-k_1-k_2-\vect{q} w_s(p)) T(k_1+k_2+\vect{q} w_s(p),k_3, k_4 - \vect{q} w_s(p) ) },\\
&{\cal T}_t^2 \equiv  {T(\vect{k_1 + \vect{q} w_t(p)},k_3,-k_1-k_3-\vect{q} w_t(p)) T(k_1+k_3+\vect{q} w_t(p),k_2, k_4 - \vect{q} w_t(p) ) }.
\end{split}
\end{equation}

It is easy to see that the same structure persists for $n$-point amplitudes. 
The key point is that the perturbative rules for the {\em integrand}
of the Witten diagram are very similar to those of flat-space Feynman diagrams
except for Bessel function factors that appear in the numerator.  However
these Bessel functions never see the BCFW deformation (By construction,
this does not change the norm of $\vect{k_1}$ and $\vect{k_n}$ which is all
the Bessel function is sensitive to). In particular, this integrand is a rational  function of $w$. Poles in the finite $w$ plane occur only when the denominator of some propagator
vanishes. Precisely when this happens, as we have emphasized above, the numerator of the 
propagator breaks up into a sum of a product of normalizable modes. These modes combine
with the other terms to make up a product of the integrand of two transition amplitudes. 

There are two other points worth noting. The first is that the mode from the propagator that enters both transition amplitudes --- on the left and the right --- is normalizable. So, even if
we start out by computing a vacuum correlator, the residues at the poles of its integrand
comprise the product of two transition amplitudes, each of which contains one normalizable 
mode. The second is that the momenta that enter these transition amplitudes depend on 
$p$ from the propagator. Therefore, the recursion relations relate a higher-point correlator 
to the {\em integrated} product of two lower-point correlators.

All that remains is to list the positions of the poles. Poles at finite $w$ are in 
one to one correspondence with all possible partitions of the momenta into two sets
with $\vect{k_1}$ in one, and $\vect{k_n}$ in the other.  In the $\phi^3$ theory
under discussion (or any theory with a polynomial interaction in $\phi$), there is also
a pole at $w = \infty$. The residue at this pole cannot be written as the product of 
lower-point transition amplitudes but must be explicitly calculated by the sum 
of all Witten diagrams where $\vect{k_1}$ and $\vect{k_n}$ meet at a point. 

Thus, for a $n$-point amplitude, we have the following recursion relations
\begin{equation}
\label{scalarecurs}
\begin{split}
&T(\vect{k_1}, \ldots \vect{k_n}) = {\cal B} + \sum_{\{\pi\},m}\int  {-i {\cal T}^2 \over 2(p^2 + \vect{K}^2)}  d p^2, \\
&{\cal T}^2 \equiv  {T(\vect{k_1}(p), \ldots \vect{k_m'}) T(-\vect{k_m'},\ldots \vect{k_n}(p) ) }.
\end{split}
\end{equation}
 The sum is over all ways of partitioning
the momenta into two sets $\{\vect{k_1},\vect{k_{\pi_2}}, \ldots \vect{k_{\pi_m}}\}$ and $\{\vect{k_{\pi_{m+1}}}, \ldots \vect{k_{n}} \}$, with $\vect{k_1}$ in one and $\vect{k_n}$ in the other. Also, 
\begin{equation}
\label{intermediatedefs}
\begin{split}
&\vect{K} = \vect{k_1} + \sum_{2}^m \vect{k_{\pi_m}};  \quad w(p)=-(\vect{K}^2 + p^2)/(2 \vect{K} \cdot \vect{q});  \quad \vect{k_1}(p)=\vect{k_1}+ \vect{q} w(p); \\  &\vect{k_n}(p) = \vect{k_n} - \vect{q} w(p); \quad \vect{k_m'} = -\vect{K} - \vect{q} w(p).
\end{split}
\end{equation}
The ``boundary term,''  ${\cal B}$, is the contribution from the pole at $w = \infty$, comprising the sum of all diagrams where $\vect{k_1}$ and $\vect{k_n}$ meet at a point. As we pointed out above, 
the mode 
corresponding to $\vect{k_m'}$ in \eqref{scalarecurs} will always be normalizable. This is implicit in \eqref{scalarecurs}.

\subsection{Gauge Fields}
\label{gaugebossec}
We now turn to the more interesting case of non-Abelian gauge fields. Our main task
is to show that, in this case, the boundary term ${\cal B}$ is zero. Hence,
the recursion relations for transition amplitudes in Yang-Mills theory form a closed set. In particular,
we can entirely bypass the computation of Witten diagrams, once we have
the three-point amplitudes in hand. We will follow, and generalize, the approach of \cite{ArkaniHamed:2008yf}.

The first step is to expand the gauge-fields about a background
\begin{equation}
{\cal A}_{\mu}^{\rm a} = A_{\mu}^{\rm a} + a_{\mu}^{\rm a},
\end{equation}
where $A_{\mu}^{\rm a}$ is the background and the fluctuations comprise $a_{\mu}^{\rm a}$. We also choose
background field gauge, so that the quadratic action for $a_{\mu}^{\rm a}$ 
is 
\begin{equation}
\label{gaugeffectact}
2 {\cal L} = D_{\mu} a_{\nu}^{\rm a} D^{\mu} a^{\nu,{\rm a}} + \left(2 F^{\mu \nu, {\rm{a}}}f^{\rm a b c} +  R^{\mu \nu} \delta^{\rm b c}\right) a_{\mu}^{\rm b} a_{\nu}^{\rm c}.
\end{equation}
We can think of our $n$-point scattering amplitude as a two-point function in this gauge --- involving momenta $\vect{k_1} + \vect{q} w$ and $\vect{k_n} - \vect{q} w$  --- and put the rest of the dynamics into the background field $A_{\mu}^{\rm a}$. The advantage of this gauge is that we can independently choose a
gauge for $A_{\mu}^{\rm a}$, which we choose using 
\begin{equation}
\label{qlightcone}
\vect{q} \cdot \vect{A}^{\rm a} = 0.
\end{equation}

If we wanted to compute bulk correlation functions, then we would also have
to work out the propagator in background field gauge. However, since we are only interested
in bulk-transition amplitudes (which are like flat-space on-shell S-matrix elements), the Ward identities \eqref{wardidentity} tell us that we can continue to use the propagator \eqref{axialpropagator}. Furthermore, we note that this
propagator itself may be written as
\begin{equation}
\label{killmomtermsprop}
\begin{split}
&G^{\text{axial},{\rm a b}}_{\mu \rho}(\vect{x}, z, \vect{x}',z') \\ &= \delta^{a b} \int {-i d^d \vect{k} d p^2 \over 2 (2 \pi)^d }   \biggl[{e^{i \vect{k} \cdot (\vect{x} - \vect{x'})}} {(z z')^{d - 2 \over 2} J_{d-2\over 2 }(p z) J_{d-2 \over 2} (p z') \left(\eta_{\mu \rho} - \delta^0_{\mu} \delta^0_{\rho}\right) \over \left(\vect{k}^2 + p^2 - i \epsilon \right)}\\
&\phantom{\delta^{a b} \int {-i d^d \vect{k} d p^2 \over 2 (2 \pi)^d }  \Bigl[}
+{\left(\partial_{\mu} \psi(x,z) - \delta_{\mu}^0 {\partial \psi(x,z) \over \partial z}\right) \left(\partial_{\rho}' \psi^*(x',z') - \delta_{\rho}^0 {\partial \psi^*(x',z') \over \partial z'}\right)  \over 
p^2 \left(\vect{k}^2 + p^2 - i \epsilon \right)}\biggr],
\end{split}
\end{equation}
where
\begin{equation}
\label{psitosubtract}
\psi(x,z) = e^{i \vect{k} \cdot \vect{x}}  z^{d-2\over 2} J_{d - 2 \over 2} (p z),
\end{equation}
and we have extended $\vect{\eta}$ to $d+1$ dimensions, by defining $\eta_{0 0} = 1$. (Recall that $0$ refers to the z-component, not the time-component.)
The Ward identities \eqref{wardidentity} now tells us that, in the computation
of transition amplitudes, we can instead use
\begin{equation}
\label{greduced}
\begin{split}
G^{\text{axial},{\rm a b}}_{\mu \rho} \sim 
 \int {-i d^d \vect{k} d p^2 \over 2 (2 \pi)^d }   \biggl[ &{ {e^{i \vect{k} \cdot (\vect{x} - \vect{x'})}}(z z')^{{d - 2 \over 2}} J_{{d - 2 \over 2}}(p z) J_{{d - 2 \over 2}} (p z') \eta_{\mu \rho}  \over \left(\vect{k}^2 + p^2 - i \epsilon \right)}\\
&+{ \delta_{\mu}^0 \delta_{\rho}^0 \left(  {\partial \psi(x,z) \over \partial z} {\partial \psi^*(x',z') \over \partial z'} - p^2 \right)  \over 
p^2 \left(\vect{k}^2 + p^2 - i \epsilon \right)}\biggr].
\end{split}
\end{equation}

Now, the form of the propagator in \eqref{greduced} makes it clear
that every propagator comes with a factor of ${1 \over w}$. On the other 
hand, factors of $w$ can appear only through derivative interactions. Our
choice of q-lightcone gauge gets rid of almost all these factors. The only
time that  we get an $\Or[w]$ vertex is in a  diagram where all the 
background fields interact with themselves and then interact with 
the fluctuating field through a single line. This line carries momentum 
$-(\vect{k_1} + \vect{k_n})$, which is orthogonal to $\vect{q}$; so, we cannot make
it obey the choice of gauge
\eqref{qlightcone}.

The reason for going through this procedure is to point out that, at large $w$,
the dominant contribution to the transition amplitude is
\begin{equation}
\label{twopointbcfw}
\begin{split}
\int &\left[A^{\mu,{\rm a}} f^{\rm a b c} \left(a_{1}^{\nu,{\rm b}} \nabla_{\mu} a_{n,\nu}^{\rm c} - a_{n}^{\nu,{\rm c}} \nabla_{\mu} a_{1 \nu}^{\rm b} \right) 
+ 2 F^{\mu \nu,{\rm a}} a_{1 \mu}^{\rm b} a_{n \nu}^{\rm c} f^{\rm a b c} + \Or[{1 \over w}] \right]  {d^{d} \vect{x} dz \over z^{d+1}}, 
\end{split}
\end{equation}
where $\vect{a_1}, \vect{a_n}$ belong to \eqref{gaugesoltime} or \eqref{gaugesolspace}.
Below,
we will suppress the color-factors, which are unimportant for our purposes.

We choose the polarization for $\vect{a_1}$ by $\vect{\ep_1} = \vect{q}$, and define $\vect{t}$ by
$ a_{1 \mu} \equiv w^{-1} \left(\partial_{\mu} \phi - t_{\mu}\right),$
where $\phi = e^{i (\vect{k_1} + \vect{q} \omega) \cdot \vect{x}} E_{{d - 2 \over 2}} (\vect{k_1},z).$
By the Ward identity, now, instead of $a_{1 \mu}$, we can use $w^{-1} t_{\mu}$ in \eqref{twopointbcfw}. 
As a result, the terms in the integrand of \eqref{twopointbcfw} die off at large $w$ if (a) $\vect{\ep_n}$ does not grow at large $w$ (which requires $\vect{\ep_n} \cdot \vect{q} = 0$) and (b) $\vect{k_1} \cdot \vect{\epsilon_n} = 0$. In $d=4$ this forces us to take $\vect{\ep_n} = \vect{q}$ also. For $d > 4$, we can choose an $\vect{\ep_n} \neq \vect{q}$ that is orthogonal to $\vect{k_1},\vect{k_n},\vect{q}$.

With this choice of $\vect{\ep_1}=\vect{q}$ and these constraints on $\vect{\ep_n}$,  we can reconstruct the integrand, up to terms that integrate to zero, 
using its poles at finite $w$. Repeating the argument above, we
get the recursion relation (using the same notation as \eqref{scalarecurs})
\begin{equation}
\label{gaugerecurs}
\begin{split}
&T(\vect{k_1},\vect{\ep_1}, \ldots \vect{k_n},\vect{\ep_n}) = \sum_{\{\pi\},m,\vect{\ep_m'}} \int   {-i {\cal T}^2 \over 2 (p^2 + \vect{K}^2)} d p^2,\\
&{\cal T}^2 \equiv {T(\vect{k_1}(p),\vect{\ep_1}, \ldots \vect{k_m'},\vect{\ep_m'}) T(-\vect{k_m'},\vect{\ep_m'}, \ldots \vect{k_n}(p),\vect{\ep_n} )}.
\end{split}
\end{equation}
This has no boundary term and the sum now also runs over all normalized polarization vectors for $\vect{k_m'}$. The definitions of \eqref{intermediatedefs} 
continue to hold.

\begin{figure}
\begin{center}
\includegraphics[width=0.5\textwidth]{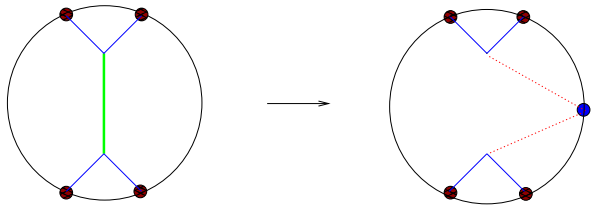}
\caption{\label{split} Recursion Relations}
\end{center}
\end{figure}

These recursion relations are shown schematically in Fig. \ref{split}. Say we set out
to compute a four-point vacuum-vacuum correlator. A typical Witten diagram
involves three ingredients: a bulk-bulk propagator shown by the heavy
line in the middle (green), four insertions of a source on the boundary shown
by the small crosshatched circles, and four bulk-boundary propagators shown
by the lines from the boundary to the bulk (blue). The recursion relations
\eqref{gaugerecurs} convert this to the integrated product of two three-point
functions. This is done by cutting open the bulk-bulk propagator and replacing
it with a product of two  normalizable modes shown by the dotted lines (red). Since these modes are normalizable the small solid circle (blue) on the boundary
is not really the insertion of a source but represents a coherent state. We need to integrate over the momentum running through these normalizable modes, which
reproduces the result of the {\em sum} of Witten diagrams.

Note that the crucial ingredient in our derivation above was the leading 
large $w$ behaviour of the integrand. In particular, we need the $\Or[w]$ 
piece, but to derive the results above, we do not need the $\Or[1]$ or $\Or[{1 \over w}]$ pieces. 
There is another route to this method that is somewhat more direct. Instead
of using the background field method, we can just do perturbation 
theory in q-lightcone gauge for the gauge-field. In this gauge, the $\vect{q} w$
momentum does not propagate in the numerator; so it is clear that every 
propagator comes with a factor of ${1 \over w}$. Except for the 
unique vertex mentioned above, there are also no interaction vertices that
are proportional to $w$ in this gauge. This immediately leads to the $\Or[w]$
term above.  
As long as one of the polarization vectors is $\vect{q}$, this analysis together with the Ward identity
\eqref{wardidentity} is enough to derive the falloff of the integrand at 
large $w$.

\subsection{Gravity}
\label{gravitysec}
We now turn to the case of graviton scattering. Say that we are considering
a n-point transition amplitude $T(\vect{k_1}, \ldots \vect{k_n})$. To analyze
the amplitude when 
we BCFW extend $\vect{k_1} \rightarrow \vect{k_1} + \vect{q} w$ and $\vect{k_n} \rightarrow \vect{k_n} - \vect{q} w$, we go to 
background field gauge where we consider the two-point function of gravitons
with these momenta in a background of soft gravitons with momenta $\vect{k_2}, \ldots \vect{k_{n-1}}$.

We have already expanded the 
gravity action about a classical background in \eqref{gravityquadratic}. However, we need
to be cautious because the background metric now also contains the fluctuations induced
by the gravitons with momenta $\vect{k_2}, \ldots \vect{k_n}$. To differentiate it from
the AdS metric, below, we will denote it by $g^b_{\mu \nu}$ ($b$ stands for background) 
and its inverse by $g_b^{\mu \nu}$.  After adding
the background gauge-fixing term, the gravity action becomes
\begin{equation}
\label{gaugefixedgravity}
S = {-1 \over 64 \pi G} \int_{\vect{x},z} \sqrt{-g_b} 
\left(g_b^{\mu \rho} g_b^{\nu \sigma} g_b^{\alpha \beta} \nabla_{\alpha} \tilde{h}_{\rho \sigma}  \nabla_{\beta} h_{\mu \nu} + 2 \tilde{h}_{\mu_1 \nu_1}  R_{\mu \rho \nu \sigma} h_{\alpha \beta} g_b^{\mu \mu_1} g_b^{\nu \nu_1} g_b^{\rho \alpha} g_b^{\beta \sigma} \right),
\end{equation}
up to boundary terms, which do not affect the bulk Green's functions and where we have written the factors of the inverse metric explicitly for reasons
that will shortly become clear. 

It is convenient to break up the background
metric into a ``pure'' AdS part and another part that comes from the 
fluctuations caused by the gravitons in our amplitude.  We write
\begin{equation}
\label{splitbackgroundmetric}
H^{\mu \nu} \equiv {g_b^{\mu \nu} - g^{\mu \nu}},
\end{equation}
where $g^{\mu \nu}$ is the inverse of the metric in \eqref{adsmetric}.  We choose q-lightcone gauge, which means that 
\begin{equation}
\label{gravqlight}
q_{\mu} H^{\mu \nu} = 0.
\end{equation}
The split in \eqref{splitbackgroundmetric} is not arbitrary since the 
AdS metric is the zero-momentum part of the background. 

Now, we work with the propagator \eqref{gravitypropagator}, which corresponds
to treating the AdS part of the quadratic action exactly but the $H^{\mu \nu}$ as a perturbation. Note that if we were to expand out \eqref{gaugefixedgravity},
in terms of $\vect{H}$ the answer would be inordinately complicated. Fortunately,
we are only interested in the $\Or[w^2]$ part of the amplitude and this is 
easy to determine. 

As above, there is a unique set of  diagrams that contribute to this action. In 
these, $H^{\mu \nu}$ carries the momentum $-\vect{k_1} - \vect{k_n} = \sum_{i=2}^{n-1} \vect{k_i}$ for which \eqref{gravqlight} cannot be chosen. The leading
$\Or[w^2]$ part of the transition amplitude then comes from
\begin{equation}
\label{leadingrav}
\int_{\vect{x},z} \sqrt{-g_b} \left[H^{\mu\nu}(-\vect{k_1}-\vect{k_n}, z) q_{\mu} q_{\nu} w^2 g^{\rho \sigma} g^{\alpha \beta} h_{\rho \alpha} (\vect{k_1} + q w, z) h_{\sigma \beta}(\vect{k_n} - q w, z)  + \Or[w] \right]. 
\end{equation}
Note that we have performed the $\vect{x}$ integral and imposed momentum conservation. So, 
the functions $\vect{H}$ and $\vect{h}$ in the expression above are written as functions
of the momenta and the radial coordinate only. 

As in the case of gauge fields above, this result may be alternately
derived by going to q-lightcone gauge from the start. Note that the terms
that appear at $\Or[w]$ and $\Or[1]$ above are significantly more complicated
than in flat-space, but we do not need their explicit forms. 

Consider the case where the polarization for $h_{\mu \nu}(\vect{k_1} + q w,z)$,
as defined in \eqref{normalsolgrav} is $\ep^1_{\mu \nu} = q_{\mu} q_{\nu}$. With
$\psi_1$ defined by 
\begin{equation}
\label{psi1tosubtract}
\psi_1(\vect{x}, z) \equiv z^{-2} E_{d \over 2}(\vect{k_1}, z) e^{i (\vect{k_1} + q w) \cdot \vect{x}},
\end{equation}
we can write (after restoring the $\vect{x}$ dependence in $\vect{h}$)
\begin{equation}
\label{simplifyh1}
\begin{split}
h_{\mu \nu}(\vect{k_1} + \vect{q} w,\vect{x},z) &= q_{\mu} q_{\nu} z^{-2} E_{d \over 2}(\vect{k}_1, z) e^{i (\vect{k_1} + \vect{q} w) \cdot \vect{x}} \\
&= {1 \over w} \left( \nabla_{(\mu} {q_{\nu)} \psi_1} - k_{1 (\mu} q_{\nu)} \psi_1 -\delta_{(\mu}^0 q_{\nu)} {\partial \psi_1 \over \partial z} - \Gamma_{\mu \nu}^{\rho} q_{\rho} \psi_1 \right) \\
&= {1 \over w} \left( \nabla_{(\mu} {q_{\nu)} \psi_1} - k_{1 (\mu} q_{\nu)} \psi_1 -\delta_{(\mu}^0 q_{\nu)} {\partial \psi_1 \over \partial z} + 2 q_{(\mu} \delta_{\nu)}^0 {\psi_1 \over z} \right) \\
&\equiv {1 \over w} \nabla_{(\mu} {q_{\nu)} \psi_1} - {1 \over w} q_{(\mu} t_{\nu)},
\end{split}
\end{equation}
where the last line defines a vector field $\vect{t}(\vect{x}, z)$, and we have used
\eqref{connectionval} to go from the first to the second line. 
However, since the dependence of $\vect{t}$ on $\vect{x}$ continues to be of the form
$e^{i (\vect{k_1} + \vect{q} w) \cdot \vect{x}}$, this itself may be further rewritten using 
\begin{equation}
\label{moresimplifyh1}
{1 \over w} q_{(\mu} t_{\nu)} = {1 \over w^2} \left( \nabla_{(\mu} t_{\nu)}  - k_{1 (\mu} t_{\nu)} -  {\delta_{(\mu}^0 \partial t_{\nu)} \over \partial z} - \Gamma^{\rho}_{\mu \nu} t_{\rho} \right) \equiv {1 \over w^2} \nabla_{(\mu} t_{\nu)}  + {1 \over w^2} t^{(2)}_{\mu \nu}(x,z),
\end{equation}
where $t^{(2)}_{\mu \nu}$ is a symmetric tensor field that is defined by the 
equation above. The Ward identity \eqref{wardidentitygravity} now tells us
that instead of using $h_{\mu \nu}(\vect{k_1} + \vect{q} w, x,z)$, we can
instead use ${1 \over w^2} t^{(2)}_{\mu \nu}(x,z)$ in \eqref{leadingrav}.

Next, note that  $h_{\mu \nu}(\vect{k_n} - \vect{q} w, x, z)$ does not itself
grow at large $w$, provided it has a polarization that is orthogonal to $\vect{q}$. So, 
we see that the following conditions are {\em sufficient}
for the integrand to behave well:
\begin{enumerate}
\item{\label{cond1}}
$\ep^n_{\mu \nu} t^{(2)}_{\rho \sigma} g^{\mu \rho}  g^{\nu \sigma} = 0$,
\item{\label{cond2}}
$\ep^n_{\mu \nu} q_{\rho}  g^{\mu \rho} = 0$.
\end{enumerate}
There is an additional possibility. If $\vect{\ep^n}$ itself has a factor of $\vect{q}$
then we can repeat \eqref{simplifyh1} to pull down another factor of ${1 \over w}$. In that
case, condition \ref{cond2} above is automatically satisfied but we can relax condition \ref{cond1}.

Hence, we see that the graviton transition amplitude is well behaved under the following 
conditions:
\begin{enumerate}
\item
$\ep^n_{i j} = q_{(i}v_{j)},$ where $\vect{v} \cdot \vect{q} = 0$ or 
\item
$\ep^n_{i j}=v^1_{(i} v^2_{j)},$ where $\vect{v^m} \cdot \vect{q} = \vect{v^m} \cdot \vect{k_1} = 0$. 
\end{enumerate}
The first set above includes the case where $\vect{v} = \vect{q}$. Also, just the requirement
that $\vect{\ep^n}$ be a valid polarization vector implies that $\vect{v} \cdot \vect{k_n} = \vect{v^1} \cdot \vect{v^2} = \vect{v^1} \cdot \vect{k_n} = \vect{v^2} \cdot \vect{k_n} = 0$.
Of course, we can interchange the role of $\vect{\ep^1}$ and $\vect{\ep^n}$ above. So, if we
take $\ep^n_{i j} = q_i q_j$, then we can have $\ep^1_{i j} = q_{(i}v_{j)},$ where $\vect{v} \cdot \vect{q} = \vect{v} \cdot \vect{k_1} = 0$ or 
$\ep^1_{i j}=v^1_{(i} v^2_{j)},$ where $\vect{v^m} \cdot \vect{q} = \vect{v^m} \cdot \vect{k_n} = \vect{v^m} \cdot \vect{k_1} = 0$.

With these conditions on the polarization vectors we find the following recursion
relations
\begin{equation}
\label{gravityrecurs}
\begin{split}
&T(\vect{k_1},\vect{\ep_1}, \ldots \vect{k_n},\vect{\ep_n}) = \sum_{\{\pi\},m,\vect{\ep_m'}} \int   {-i {\cal T}^2 \over 2 (p^2 + \vect{K}^2)} d p^2,\\
&{\cal T}^2 \equiv {T(\vect{k_1}(p),\vect{\ep_1}, \ldots \vect{k_m'},\vect{\ep_m'}) T(-\vect{k_m'},\vect{\ep_m'}, \ldots \vect{k_n}(p),\vect{\ep_n} )}.
\end{split}
\end{equation}
The sum over $\vect{k_m'}$ is again over all valid polarization vectors for this momentum and 
the other quantities in \eqref{gravityrecurs} are defined the same way as in \eqref{gaugerecurs} and \eqref{scalarecurs}. In particular, the definitions of
\eqref{intermediatedefs} continue to apply.
These recursion relations are structurally identical to \eqref{gaugerecurs} although we should
remember that each polarization vector, $\vect{\ep_m}$, above is a two-index tensor in \eqref{gravityrecurs},
and only a vector in \eqref{gaugerecurs}.

\subsection{Polarization Vectors \label{polvecsec}}
As we have pointed out above, given a transition amplitude with
some external polarization vectors, we cannot typically compute it 
by means of a single BCFW extension. However, since the amplitude
depends {\em linearly} on the polarization vectors, we can often decompose
it into a sum over amplitudes, each of which behaves well under some BCFW
extension. 

To see this more clearly, say that we are interested in a correlator 
of operators $O^I$ where the capital $I$ index indicates that these operators
transform in some representation of $SO(d-1,1)$. A n-point correlator of 
these operators can be written as
\begin{equation}
\label{lorentzstructure}
 T^{I_1 \ldots I_n}(\vect{k_1}, \ldots \vect{k_n}) \equiv \langle O^{I_1}(\vect{k_1}) \ldots O^{I_n}(\vect{k_n}) \rangle,
\end{equation}
where $T^{I_1 \ldots I_n}$ is some tensor of the Lorentz 
group. Computing the correlator is the same as computing all components
of  $\vect{T}$.

There are further constraints on the \eqref{lorentzstructure}. 
For example, it must be invariant under a simultaneous interchange of $\vect{k_m}$ and $\vect{k_n}$, and $I_m$ and $I_n$. Second, $\vect{T}$ must be conserved
with respect to each momentum when we are dealing with
conserved currents or the stress tensor. 
 There are many additional constraints \cite{Osborn:1993cr} that come from conformal
invariance, which strongly restrict the form of
$\vect{T}$ in  \eqref{lorentzstructure} for small $n$. However, these are easier to see in position
space, and we will not use them in this section.

If we dot \eqref{lorentzstructure} with some external polarization vectors, $\vect{\ep^1}, \ldots \vect{\ep^n}$, 
we find
\begin{equation}
\label{tdottedwithep}
\langle O^{I_1}(\vect{k_1}) \ldots O^{I_n}(\vect{k_n}) \epsilon^1_{I_1} \ldots \epsilon^n_{I_n}  \rangle =  T^{I_1 \ldots I_n}(\vect{k_1}, \ldots \vect{k_n}) \epsilon^1_{I_1} \ldots \epsilon^n_{I_n}.
\end{equation}
If we had some means of computing the left hand side for arbitrary polarizations, this would enable us to extract all the components of  $\vect{T}$. However,
the BCFW extension that we have described above puts constraints on the
polarization vectors. Can we recover information about all components
by combining different pairs of BCFW extensions?  

We will consider the four-point function in detail below. As we mentioned above,
 there are dynamical constraints on this correlator; however, we will only
focus on the fact that it must be conserved. This is because if we can show, without using any other dynamical information,
that all possible components of $\vect{T}$ in \eqref{lorentzstructure} in a  four-point function are accessible by our methods, then this 
would be sufficient to show that arbitrary n-point correlators are also
accessible.  

To see this, consider what happens if we add a fifth index to \eqref{lorentzstructure}. The recursion relations \eqref{gaugerecurs} and \eqref{gravityrecurs} place constraints only on pairs of polarization vectors. So, if
all possible four-point tensor structures are calculable by choosing 
and extending pairs from  $\vect{k_1} \ldots \vect{k_4}$, then we can choose the 
polarization vector for $\vect{k_5}$ arbitrarily. Moreover, every allowed
conserved five-index tensor can be written as
\begin{equation}
\label{t5decomposedin4}
T^{I_1 \ldots I_5}(\vect{k_1}, \ldots \vect{k_5}) = \sum_{\alpha} T_{\alpha}^{I_1 \ldots I_4}(\vect{k_1}, \ldots \vect{k_5}) v_{\alpha}^{I_5}(\vect{k_1}, \ldots \vect{k_5}),
\end{equation}
where $v_{\alpha}$ is a list of vector functions, which are orthogonal to $\vect{k_5}$, indexed by $\alpha$. Evidently, if we can
choose the polarization vector $\vect{\epsilon^5}$ arbitrarily and we can 
also access all tensors $T_{\alpha}^{I_1 \ldots I_4}$, then we can also
access any five-index tensor.

While the computability of all four-point tensor structures is sufficient
to ensure the computability of higher-point structures, it is not necessary. 
In some cases (such as gravity in $d=5$), as we will see below, it is not possible to compute all possible tensor-structures for the four-point correlator. 
However, adding additional particles now gives us more choices of BCFW 
extensions and, for a sufficient number of external particles, we can once
again access all allowed tensor structures.

\subsubsection{Polarizations for Gauge Bosons}
Let us start with the case of $d=4$. For each momentum, $\vect{k_m}$, there
are three allowed polarization vectors. So, a four-point correlator
$\vect{T}$ has, a priori, $3^4 = 81$ possible components. On the other hand, 
given a pair of momenta, say  $\vect{k_1}$ and $\vect{k_4}$, there are two choices of $\vect{q}$, which we denote by $\vect{q^1}$ and $\vect{q^2}$.
The constraints worked out in subsection \ref{gaugebossec} now tell us 
that we either take $\vect{\ep^1} = \vect{q^1}, \vect{\ep^4} = \vect{q^1}$ or $\vect{\ep^1} = \vect{q^2}, \vect{\ep^4} = \vect{q^2}$.
The polarization vectors for $\vect{k_2}$ and $\vect{k_3}$ can be arbitrary
and there are $3 \times 3 = 9$ combinations of these.
So, extending momenta $\vect{k_1}$ and $\vect{k_4}$ should allow us
to compute $2 \times 9 = 18$ components of $\vect{T}$.

We can choose 6 distinct pairs with 4 particles. So, by making all possible
BCFW extensions we get $18 \times 6 = 108$ pieces of data about the 
components of $\vect{T}$. One would naively think that this forms an 
overcomplete basis for the $81$ numbers that we wish to extract. However,
this is a little too quick. This is because, if $\vect{T}$ is of 
the form
\begin{equation}
\label{antisymmetric}
T^{i_1 i_2 i_3 i_4} = \epsilon^{j_1 j_2 j_3 j_4}\left(\delta_{j_1}^{i_1} - {k_{1 j_1} k_1^{i_1} \over \vect{k_1}^2} \right)\left(\delta_{j_2}^{i_2} - {k_{2 j_2} k_2^{i_2} \over \vect{k_2}^2} \right) \left(\delta_{j_3}^{i_3} - {k_{3 j_3} k_3^{i_3} \over \vect{k_3}^2} \right) \left(\delta_{j_4}^{i_4} - {k_{4 j_4} k_4^{i_4} \over \vect{k_4}^2} \right),
\end{equation}
it will be invisible to all our choices of BCFW extension. So, we can compute
only 80 components of $\vect{T}$ using all 6 BCFW extensions. Fortunately,
this is not a problem since the choice of $\vect{T}$ in \eqref{antisymmetric} is completely
antisymmetric under the simultaneous exchange of any pair of indices
and momenta. For example, interchanging $(\vect{k_1}, i_1)$ and
$(\vect{k_2}, i_2)$ changes the sign of \eqref{antisymmetric}. Consequently, this tensor
structure is inconsistent with the symmetries of the correlator and is not
allowed. So, we can compute all possible tensor structures in correlators
of conserved currents, in $d=4$, by means of the BCFW extension.

We now turn to $d=5$. Each momentum has $4$ possible associated polarization
vectors; so $\vect{T}$ has, a priori,  a total of $4^4 = 256$ components for a four-point
correlator. Given a pair of momenta, say  $\vect{k_1}$ and $\vect{k_4}$,
we now have a continuous family of choices for $\vect{q}$. Moreover, if we
choose $\vect{\ep^1} = \vect{q}$, we can also take $\vect{\ep^4} = \vect{v^4}$, where $\vect{v^4}$ is a vector
orthogonal to $\vect{q}$ and $\vect{k_4}$. Similarly, by taking $\vect{\ep^4} = \vect{q}$, we can 
take $\vect{\ep^1} = \vect{v^1}$, orthogonal to $\vect{q}$ and $\vect{k_1}$. The polarizations for 
$\vect{k_2}$ and $\vect{k_3}$ can, again, be arbitrary. Focusing, momentarily,
only on the indices corresponding to $i_1$ and $i_4$, a general 2-index 
 conserved tensor can be written as
\begin{equation}
\label{gaugetdecomp}
T^{i_1 i_4} = T_{\text{tl}}^{i_1 i_4} + A \eta^{j_1 j_4} \left(\delta^{i_1}_{j_1} - {k_1^{i_1} k_{1 j_1} \over \vect{k_1}^2} \right) \left(\delta^{i_4}_{j_4} - {k_4^{i_4} k_{4 j_4} \over \vect{k_4}^2} \right),
\end{equation}
where $A$ is some constant and $\vect{T_{\text{tl}}}$ is (a) traceless and (b) conserved with respect
to both $\vect{k_1}$ and $\vect{k_4}$. The polarization-combinations above allow us to detect all components of $\vect{T}_{\text{tl}}$ but are insensitive to the presence of $A$. 

Since $\vect{T_{\text{tl}}}$ has 8 components, making all possible choices
of polarizations for $\vect{k_2}$ and $\vect{k_3}$ gives us $8 \times 16 = 108$
pieces of data. Now, making all 6 possible BCFW extensions gives us $6 \times 108 = 648$ pieces of data. Also, for gauge bosons in $d=5$, there is no inaccessible tensor like \eqref{antisymmetric}.  So, by making
all possible BCFW extensions, we obtain an overcomplete set of equations for 
the $256$ allowed components of $\vect{T}$ that can all, hence, be determined.

The calculation for $d \geq 6$ goes as follows. There are $d-1$ possible
polarizations for each momentum leading to $(d-1)^4$ independent components. 
Given a pair $\vect{k_1}, \vect{k_4}$ we now not only have a $d-4$ parameter
continuous family of choices for $\vect{q}$, we can 
also choose $\vect{v^1}$ and 
$\vect{v^4}$ in several different ways provided we keep them orthogonal to $\vect{q}$. Hence, just as above, in the $i_1, i_4$ space 
we can detect all tensors that are conserved with respect to $\vect{k_1}$
and $\vect{k_4}$ and are traceless. There are $(d-2)^2-1$ linearly independent
tensors of this sort. So, one BCFW extension gives us 
$(d-1)^2 \left[(d-2)^2 - 1 \right]$
pieces of data. The six possible BCFW extensions give us $6 (d-1)^2 \left[(d-2)^2 - 1\right]$ pieces of data that form an overcomplete set for the $(d-1)^4$ 
independent allowed components of $\vect{T}$ in $d$-dimensions.\footnote{In fact, we have
already seen that ${d \choose 4}$ of these components must be zero. However, we have to put
this in from the start only for $d=4$; in higher $d$, the recursion relations will
allow us to compute these components and verify that they vanish.}

\subsubsection{Polarizations for Gravitons}
We now turn to the case of gravity. We start with $d=4$ and then go on to 
higher $d$. As we will see below this analysis has similarities but also
important differences with the analysis for gauge bosons.

Polarization vectors for a graviton with momentum $\vect{k}$ are given by symmetric traceless tensors that are orthogonal to $\vect{k}$. In $d=4$, this implies
that we have 5 possible polarizations for each momentum. As we pointed out above, if we extend $\vect{k_1}$ and
$\vect{k_4}$  there are two possible choices of $\vect{q}$, which we denote by
$\vect{q^1}$ and $\vect{q^2}$. If we choose the polarization vector for
particle $\vect{k_1}$ by $\ep^1_{i j} = q_i q_j$, (where $\vect{q}$ is either of the
two allowed choices) then we must choose
 $\ep^4_{i j} = q_i q_j$ or $\ep^4_{i j} = q_{(i} v^4_{j)}$, where, as above, $\vect{v^4}$ is a 
vector orthogonal to $\vect{q}$ and $\vect{k_4}$.  The choice of 
 $\ep^4_{i j} = q_i q_j$ also behaves well under the BCFW extension if we 
choose $\ep^1_{i j} = q_{(i} v^1_{j)}$, where $\vect{v^1}$ is orthogonal to $\vect{q}$ and $\vect{k_1}$. The polarizations for $\vect{k_2}$
and $\vect{k_3}$ can be arbitrary.  

So, extending $\vect{k_1}$ and $\vect{k_4}$
by $\vect{q^1}$ allows us to calculate $3 \times 25 = 75$ combinations of polarizations. Extending $\vect{k_1}$ and $\vect{k_4}$ by $\vect{q^2}$ allows us to calculate
another 75. So, with all 6 possible BCFW extensions we can compute
$6 \times (75 + 75) = 900$ polarization-combinations. Naively, this would
seem to give an overcomplete set for the 625 distinct components of $\vect{T}$.

However, as above, this is not quite correct. In particular, consider 
a $\vect{T}$ of the form 
\begin{equation}
\label{anomaly}
T^{i_1 j_1 \ldots i_4 j_4} = S_{1 2 3 4} \left[\ep^{i_1 \ldots i_4} \ep^{j_1 \ldots j_4}\right],
\end{equation}
where the operator $S_{1 2 3 4}$ symmetrizes its argument in  $(i_1,j_1),(i_2,j_2), (i_3,j_3), (i_4,j_4)$, makes it traceless in these pairs, and projects the $(i_m,j_m)$ component
orthogonal to $\vect{k_m}$. The $\vect{T}$ in \eqref{anomaly} cannot be detected
by any of the 900 BCFW extensions described above. Moreover, unlike in the 
case of gauge bosons, this tensor is symmetric under the simultaneous 
interchange of Lorentz indices and momenta. For example, if we interchange
$(i_1,j_1)$ and $(i_4,j_4)$, and also interchange $\vect{k_1}$ and $\vect{k_4}$
then $\vect{T}$ returns to itself, because each $\epsilon$ tensor contributes
a factor of $(-1)$ leading to an overall factor of unity. We are also
not aware of any other argument that would allow us to exclude this term.
So, more precisely,
the BCFW procedure allows us to compute 624 out of the 625 components
of a general $\vect{T}$. 

It would be interesting to understand the coefficient of this term
for the four-point function of stress-tensors given Einstein gravity in AdS$_5$. Also, it would be nice if conformal
invariance or other restrictions allowed us to fix this term; the
studies of \cite{Drummond:2006by,Eden:2000bk} might be useful in this
context.  

Notice that this problem does not occur for five- and higher-point amplitudes
in $d=4$. There is no tensor, in four dimensions, that is completely antisymmetric
in five indices. So, for higher-point amplitudes it seems possible to compute
any polarization-combination using the BCFW extension.

We now turn to $d=5$. Each momentum has $9$ possible associated polarization
vectors; hence, $\vect{T}$ has a total of $9^4 = 729$ components for a four-point
correlator. The important difference from $d=4$ is that given a pair of momenta, say  $\vect{k_1}$ and $\vect{k_4}$,
we now have a continuous family of choices for $\vect{q}$. Once again, if 
we choose $\ep^1_{i j} = q_i q_j$, we can take $\vect{\ep^4} = \vect{\ep^1}$ or  $\ep^4_{i j} = q_{(i} v^4_{j)}$. In $d=6$ and higher, as we will
see below, we have additional choices for $\vect{\ep^4}$ that are not available here. This is because in $d=6$ (or higher), there are two (or more) linearly
independent vectors that
are orthogonal to $\vect{q}$, $\vect{k_1}$ and $\vect{k_4}$.

Let us focus, momentarily,
only on the indices corresponding to $\vect{k_1}$, which are  $i_1, j_1$, 
and the indices corresponding  $\vect{k_4}$, which are $i_4, j_4$. For
the sake of simplicity in this analysis, we also assume (without loss of 
generality for our purposes) that $\vect{k_1}$
is orthogonal to $\vect{k_4}$.  

First, we consider the number of linearly independent polarization-combinations
we can obtain by taking $\vect{\ep}^1_{i j} = q_i q_j$ and $\vect{\ep}^4_{i j}
= q_i q_j$ or $\vect{\ep}^4_{i j} = q_i v^4_j$ and let us, for the sake 
of illustration,
also choose $\vect{v_4}$ to be orthogonal to all three $\vect{k_1}, \vect{k_4}$
and $\vect{q}$. If the indices $i_1, j_1, i_4, j_4$ run only over the three-dimensional subspace orthogonal to $\vect{k_1}$ and $\vect{k_4}$, then given a tensor
that is symmetric and traceless in each of $(i_1,j_1)$ and $(i_4,j_4)$ we
can write
\begin{equation}
\label{so3decomp}
T^{i_1 j_1 i_4 j_4} = T_{(4)}^{i_1 i_4 j_1 j_4} + \epsilon^{i_1 i_4 l} T_{(3)}^{l j_1 j_4} +  \delta^{i_1 i_4} T_{(2)}^{j_1 j_4} + \delta^{i_1 i_4} T_{(1)}^{j_1 j_4} + T_{(0)} \delta^{i_1 i_4} \delta^{j_1 j_4} + (i_1 \leftrightarrow j_1) + (i_4 \leftrightarrow j_4).
\end{equation}
Here the subscript under each tensor indicates which representation of $SO(3)$
it belongs to. (The $SO(3)$ is the group of rotations in the three-dimensional space
orthogonal to $\vect{k_1}$ and $\vect{k_4}$.) So $\vect{T_{(4)}}, \vect{T_{(3)}}, \vect{T_{(2)}}$ are completely symmetric and traceless in any pair of indices and  $\vect{T_{(1)}}$ is antisymmetric. 

By just using $\ep^1_{i j} = q_i q_j$ and $\ep^4_{i j} = q_i q_j$, we can 
completely extract the $\vect{T_{(4)}}$ term in \eqref{so3decomp}. To see this,
note that, in this orthogonal 3 dimensional subspace, we can choose $q$ as 
\begin{equation}
\label{generalqchoice}
q = (1, {x^2 - 1 \over 2 i x}, {x^2 + 1 \over 2 x}).
\end{equation}
Dotting $\vect{T}$ with four-copies of $q$ we find 
\begin{equation}
T^{i_1 j_1 i_4 j_4} q_{i_1} q_{j_1} q_{i_4} q_{j_4} = T_{(4)}^{i_1 j_1 i_4 j_4} q_{i_1} q_{j_1} q_{i_4} q_{j_4}.
\end{equation}
We can extract the coefficients of $x^4, x^3, \ldots x^{-3}, x^{-4}$ in this
dot product by allowing $x$ to run over various values. More precisely, 
we could take $x = e^{i \theta}$ and Fourier transform in $\theta$
that would give us these 9 different coefficients. These correspond exactly
to the 9 different components of $\vect{T_{(4)}}$. Now, consider taking $\vect{v^4}$ to be orthogonal to both $\vect{k_1}$ and
$\vect{k_4}$. If we now take $\ep^4_{i j} = q_{(i} v^4_{j)}$, with $\vect{v^4}$
orthogonal to {\em both} $\vect{k_1}$ and $\vect{k_4}$ then 
we can also extract the $\vect{T_{(3)}}$ term in \eqref{so3decomp}. This gives us
7 more terms. 

So far we have considered tensors that are orthogonal to both $\vect{k_1}$
and $\vect{k_4}$. However, we are also allowed tensors of the form
\[
(T')_{(3)}^{i_1 j_1 i_4 1} + (T')_{(3)}^{i_1 j_1 1 i_4}
\]
This tensor has one leg along the direction of $\vect{k_1}$, which is 
indicated by the index $1$ in the superscript,   and the other indices
$i_1,j_1,i_4$ run in the space orthogonal to both $\vect{k_1}$ and $\vect{k_4}$.  We have chosen this tensor to be completely symmetric in these three-indices 
and placed a prime-symbol on it to distinguish it from the $\vect{T_{(3)}}$ that appeared above. This is an allowed tensor structure because it is conserved (recall that we have taken $\vect{k_1} \cdot \vect{k_4} = 0$) and, by construction, it is symmetric in
its last two indices.   By taking $\vect{v^4} = \vect{k_1}$  We can 
now extract the 7 components of this tensor. 

If we also take $\ep^4_{i j} = q_i q_j$ and allow $\ep^1_{i j} = q_i v^1_j$,
then we get a total of $9 + 7 \times 4 = 37$ components. Combining the $81$
polarizations from $\ep^2, \ep^3$ we get $37 \times 81 = 2997$ components.
The six different BCFW extensions lead us to $2997 \times 6 = 17982$ 
different polarization-combinations, which would seem to 
form a highly overcomplete basis
for the $9^4 = 6561$ distinct possibilities. 

However, we need to be careful. As above, any tensor 
of the form (all indices now again run over all 5 dimensions)
\begin{equation}
\label{anomaly54}
T^{i_1 j_1 \ldots i_4 j_4} = S_{1 2 3 4} \left[A_1^{i_1 \ldots i_4} A_2^{j_1 \ldots j_4}\right],
\end{equation}
where $A_1, A_2$ are tensors that are antisymmetric
in any interchange of indices, cannot be detected by BCFW extensions. 

Even with
a five-point correlator, we cannot detect the part of the correlation function
that is proportional to 
\begin{equation}
\label{anomaly55}
T^{i_1 j_1 \ldots i_5 j_5} = S_{1 2 3 4 5} \left[\epsilon^{i_1 \ldots i_5} \epsilon^{j_1 \ldots j_5}\right].
\end{equation} 
So, in this case, we can get only $9^5 - 1$ out of the $9^5$ distinct 
polarization combinations. For six- and higher-point correlators, there are
no completely antisymmetric tensors and we can get all possible polarizations
using the BCFW extension.

We will not explicitly work out the combinatorics for $d=6$ and higher
since they are very similar to the calculations above. One important 
distinction is if we choose $\ep^1_{i j} = q_{i} q_{j}$, we can choose
$\ep^4_{i j} = v^{4 1}_{(i} v^{4 2}_{j)}$ where $\vect{v^{41}} \cdot \vect{v^{4 2}} = 0$ and both these vectors are orthogonal to $q, \vect{k_1}, \vect{k_4}$. This allows us to also detect tensors of the form \eqref{anomaly54} or \eqref{anomaly55}. So, for $d=6$ or higher, we can compute all polarization-combinations
using the BCFW extension. 

\section{Supersymmetric Theories \label{secsusytheories}}
We now turn to the generalization of these recursion relations to theories
with supersymmetry. In this section we will use the easily derived fact 
that as long as the particles that we are BCFW extending are gravitons
or gauge bosons, the behaviour of the amplitude at $w \rightarrow \infty$
is not affected by the presence of additional matter particles. Moreover,  \eqref{gaugerecurs} and \eqref{gravityrecurs} continue to hold
for Yang-Mills theory and gravity coupled to matter with the modification
that the sum over polarizations must be expanded to run over these particles
as well.

The basic idea here is the same as the one used in the extension of 
the BCFW-recursion relations to supersymmetric theories in flat-space \cite{Brandhuber:2008pf,ArkaniHamed:2008gz,Lal:2009gn}. As we have
pointed out above, the behaviour under BCFW extension of amplitudes with external gravitons or gauge bosons is better than that of amplitudes with external
scalars or fermions. In fact, if we BCFW-extend external scalars, our recursion relations
would involve an unwieldy boundary term that would need to be calculated explicitly
using Witten diagrams.\footnote{We should mention here that the physical intuition 
that underlies this paper: the fact  that BCFW-deformed amplitudes
are dominated by interactions at a small number of points, 
 was also used  to study these amplitudes in Yang-Mills theory coupled to matter. If we
consider a single BCFW-extended matter-line interacting with gauge bosons, the powers of 
$w$ in the amplitude are correlated with the number of color-generators that appear. This leads to some surprising cancellations
in gauge theories with matter at one-loop \cite{Lal:2009gn,Lal:2010qq}.}

In theories with supersymmetry, however, we can relate amplitudes involving
external matter particles to amplitudes involving gluons or gravitons. 
In flat-space, in ${\cn =4}$ SYM  (${\cn = 8}$ SUGRA), we can always
convert at least two particles in a scattering amplitude to 
negative helicity gluons (gravitons) \cite{Brandhuber:2008pf,ArkaniHamed:2008gz}.
In theories with less supersymmetry, such as ${\cn =2}$ SYM, we can still convert
two particles, either to negative helicity gluons or to positive
helicity gluons \cite{Lal:2009gn}.

In AdS, the situation is somewhat different. In $d=4$, for example, the
constraints on the polarization vectors for gauge bosons that we have enumerated above can be summarized by stating that for the amplitude to behave well,
it is necessary for both BCFW-extended particles to have the {\em same}
polarization vector. 
However, even maximal supersymmetry
does {\em not} always allow us to transform two particles to gluons or gravitons with the same polarization vector.  Consequently, we cannot compute correlation functions
involving arbitrary operators in the same multiplet as a conserved current
or the stress tensor. However, it is possible to compute a subset of 
correlators.

Before we show this in detail, we remind the reader that 
\ads[d+1] supergroups do not 
exist for $d > 6$ and we are interested in the cases $d=4,5,6$ \cite{Nahm:1977tg}. The reader may be more comfortable thinking about superconformal algebras in flat-space. However, although this algebra is very powerful, since we are working in momentum space, we will only use its super-Poincare subgroup.

\subsection{Supersymmetric Theories with $d = 4$}
We start by examining the case for $d=4$. For simplicity, and consistency
with standard notation, we consider a Euclidean metric on the 
boundary in this section. In $d=4$, the 
superconformal group is $SU(2,2|{\cal N})$. For ${\cal N} = 4$, we have the 16 supercharges $Q^I_{\alpha},\bar{Q}_{\dot{\alpha} I}$ and their conformal 
partners  $S^I_{\alpha},\bar{S}_{\dot{\alpha} I}$. (We follow the conventions of  \cite{Kinney:2005ej}, so $I$ is an R-symmetry
index and $\alpha, \dot{\alpha}$ are spacetime spinor indices; see also \cite{Dolan:2002zh}.) We will not make any use of the $S^I_{\alpha},\bar{S}_{\dot{\alpha} I}$ supercharges at all here.

The stress tensor multiplet for ${\cal N} = 4$ is enumerated in \cite{Gunaydin:1984fk,Dolan:2002zh} and, for the reader's convenience, we list its state content in Table \ref{stresstensord4}. The charges that we have tabulated are the charges under dilatations,  which form a SO(2) subgroup, rotations of the boundary coordinates, which 
constitute a SO(4) subgroup, and the R-symmetry SU(4). A number in brackets
next to the dilatation charge indicates the multiplicity. A minus sign
indicates that the state-content of this  representation must be subtracted off
from the state-content of the positive representations. This occurs
when some conformal representation becomes short. If we construct conformal
representations through a field theory, these negative states are indicative of the
equations of motion like, for example, the fact that the stress tensor must be conserved. (See \cite{Barabanschikov:2005ri} for other examples.) The stress-tensor itself is the representation
with a dilatation charge of $4$ that is invariant under the R-symmetry
and transforms as a $(1,1)$ under the SO(4).

\begin{table}[!h]
\begin{center}
\begin{tabular}{|c|c|c|}
	\hline         
$SO(2)$ & $SO(4)$ & $SU(4)$ \\ \hline
2 & 0,0 &0,2,0\\
5/2 & 0,1/2 & 1,1,0\\5/2 & 1/2,0 & 0,1,1\\
3 & 0,0 & 0,0,2 \\3 & 0,0 & 2,0,0\\ 3 & 0,1 & 0,1,0 \\ 3 & 1/2,1/2 & 1,0,1\\
3 & 1,0 & 0,1,0 \\ 7/2 & 0,1/2 & 1,0,0 \\ 7/2 & 1/2,0 & 0,0,1\\7/2 & 1/2,1 & 0,0,1
\\7/2 & 1,1/2 & 1,0,0 \\ 
(2) 4 & 0,0 & 0,0,0\\ 4 & 1,1 & 0,0,0\\(-)4 & 0,0 & 1,0,1 \\(-)9/2 & 0,1/2 & 0,0,1 \\ (-)9/2 & 1/2,0 & 1,0,0\\ (-)5 & 1/2,1/2 & 0,0,0	\\\hline
\end{tabular}
\caption{Stress Tensor Multiplet in $d=4$ with ${\cn = 4}$: Conformal Representations \label{stresstensord4} }
\end{center}
\end{table}

We are actually interested in the transformation properties of 
the states in this representation under the super-Poincare group. So, we
discard the SO(2) information and reduce the SO(4) representations
under the little group SO(3). This procedure leads to the representations 
listed in Table \ref{stresstensoreduced}. There are no negative states in this picture because
the equations of motion, such as the conservation of the stress tensor, are built in from
the start.
\begin{table}
\begin{center}
\begin{tabular}{|c|c|}\hline         
$SO(3)$ & $SU(4)$ \\ \hline
(2) 0&0,0,0\\
0&0,0,2\\
0&0,2,0\\
0&2,0,0\\
1/2&0,0,1\\
1/2&0,1,1\\
1/2&1,0,0\\
1/2&1,1,0\\
(2) 1&0,1,0\\
1&1,0,1\\
3/2&0,0,1\\
3/2&1,0,0\\
2&0,0,0 \\ \hline
\end{tabular}
\caption{Stress Tensor Multiplet in $d=4$ with ${\cn = 4}$: Poincare representations \label{stresstensoreduced}}
\end{center}
\end{table}

They key point, and the reason for doing this, is that the resultant representations of the Poincare group are all obtainable
by starting with the lowest-helicity state of the stress tensor (which has helicity $-2$)
and acting on it with the positive helicity supercharges. 

We now state this more formally. Say that we wish to BCFW extend the momenta $\vect{k_1}$ and $\vect{k_n}$. We can choose two linearly independent null vectors to span the two dimensional vector space that is spanned by these momenta. We choose these vectors to be
\begin{equation}
\label{twonullvecs}
\vect{n_1} = \vect{\lambda_1 \bar{\lambda}_1}; \quad \vect{n_2}= \vect{\lambda_2 \bar{\lambda}_2},
\end{equation}
where $\vect{\lambda_m}$ and $\vect{\bar{\lambda}_m}$ are spinors \cite{Dixon:1996wi}. In particular,
we have $\vect{k_1} = a_{1 1} \vect{n_1} + a_{1 2} \vect{n_2}$ and $\vect{k_n} = a_{n 1} \vect{n_1} + a_{n 2} \vect{n_2}$ where the $a$ are some coefficients.
We also choose the vector $\vect{q}$ by 
\begin{equation}
\label{qchoice1}
q_{\alpha \dot{\alpha}}= \la_{1 \alpha} \lb_{2 \dot{\alpha}}.
\end{equation}
This is clearly null and orthogonal to $\vect{k_1}$ and $\vect{k_n}$. 

Next, we assemble the vector of $2 {\cal N}$-supercharges:  ${\cal Q}_{+}^A=\{\dotl[Q^I, \la_{2}], \dotlb[\bar{Q}_I, \lb_1] \}$. $A$ runs over $1 \ldots 2 {\cal N}$ because there are
${\cal N}$ distinct  $Q$-supercharges and also ${\cal N}$ distinct $\bar{Q}$-supercharges
in this list. These are our ``positive helicity'' supercharges. The component of
the stress tensor with maximally negative helicity is 
\begin{equation}
\label{neghelcomp}
 T_{--}(\vect{k_1})=T_{i_1 j_1}(\vect{k_1}) q_{i_2} q_{j_2} \eta^{i_1 i_2} \eta^{j_1 j_2},
\end{equation}
with an analogous definition for $T_{--}(\vect{k_n})$.
 
We can generate all states in the stress-tensor multiplet given in Table \ref{stresstensoreduced} by acting on this operator with all possible combinations of the supercharges. More precisely,
given a list of Grassmann parameters $\eta_A$, where $A$ again runs from $1 \ldots 2 {\cal N}$,
we then construct the two functions (with $m=1$ or $m=n$)
\begin{equation}
\label{tcoherent}
T_m(\eta) = U_{+}(\eta) T_{--}(\vect{k_m}) U_{+}(-\eta); \quad U_{+}(\eta) \equiv e^{{\cal Q}^A_{+} \eta_{A}}. 
\end{equation}

The expansion of these operators in the $2 {\cal N}$ Grassmann parameters $\eta_{A}$
contains all the original operators listed in Table \ref{stresstensoreduced}.  With ${\cal N} = 2$, a similar expression exists for operators in
the same multiplet as a conserved current.

 We pause to note that the existence of a form like \eqref{tcoherent} for the 
operators in a representation implies 
that the representation is half-Bogomol'nyi-Prasad-Sommerfeld (BPS). This is because
every state in the representation is annihilated by half of the supercharges; for example, the highest weight state is annihilated by all the supercharges
of ``negative'' helicity.
Of course, not all half-BPS representations have this form. 
Fortunately, in $d=4,5,6$ every half-BPS representation that contains a stress-tensor
or a conserved current can be written in this form.

Now, consider a $n$-point correlator that involves two operators from \eqref{tcoherent} with
the {\em same} Grassmann parameter and $n-2$  other operators, which we denote below by the composite operator $O_{\mathcal C}$. The fact that this correlator is
invariant under supersymmetry transformations implies
\begin{equation}
\label{alltostress}
\langle T_1(\eta) T_n(\eta) O_{\mathcal C}  \rangle = \langle T_{--}(\vect{k_1}) T_{--}(\vect{k_n}) O^{\prime}_{\mathcal C} \rangle,
\end{equation}
where $O^{\prime}_{\mathcal C} \equiv U_{+}(-\eta) O_{\mathcal C} U_{+}(\eta)$. The right hand side 
can be computed by
BCFW recursion as explained above.

So, 
supersymmetry allows us to compute a ``diagonal'' subset of correlators i.e correlators of operators in the stress-tensor multiplet where
at least two Grassmann parameters are the same. 

In our analysis above, we could have made a different choice of $\vect{q}$. This choice is given by  
\begin{equation}
\label{otherqchoice}
(q_2)_{\alpha \dot{\alpha}} =  \la_{2 \alpha} \lb_{1 \dot{\alpha}}.
\end{equation}
This is also null and orthogonal to $\vect{k_1}$ and $\vect{k_n}$. Under rotations in the  subspace orthogonal to $\vect{k_1}$ and $\vect{k_n}$, 
the transformation of this vector $\vect{q_2}$ is opposite to the transformation of the vector $\vect{q}$. When we used $\vect{q}$ as a polarization 
vector, we referred to it as having ``negative helicity.'' In this terminology, $\vect{q_2}$ has ``positive helicity.'' Instead of building all states in the stress-tensor multiplet
by acting with positive-helicity supercharges on the state with maximally-negative helicity,
we could instead build these states by acting on the state with maximally-positive helicity
with the negative-helicity supercharges. This analysis allows us to compute a correlator
of operators in the same multiplet as the stress tensor using the other possible
BCFW extension, which depends on $\vect{q_2}$ above.  This 
 gives us another diagonal subset of correlators. However, the union of these two diagonal subsets is still smaller than the full
set of correlators. 

In certain cases, superconformal symmetry places strong constraints on 
correlation functions. For example, the full four-point correlator
in ${\cn=4}$ SYM, 
 can be reduced to one independent function in position space \cite{Eden:2000bk,Drummond:2006by}. So, in principle, even the diagonal subset above
should give enough information to entirely determine the correlator. It 
would be nice to translate this analysis of constraints to momentum space and see
this explicitly.
But, in general, we would like to compute correlators where all Grassmann parameters are arbitrary.
This is possible with flat-space amplitudes; the difficulty here 
is that we have stricter
constraints on the polarization-combinations that behave well 
under the BCFW extension.

\subsection{Supersymmetric Theories with d = 6}
We now turn to $d=6$. The bosonic subgroup of the $d=6$
superalgebra is $SO(6,2) \otimes Sp(2 {\cal N})$ that has maximal compact
subgroup $SO(2) \otimes SO(6) \otimes Sp(2 {\cal N})$. The supercharges live in a 6 dimensional chiral-spinor representation (with eigenvalues ${\pm 1/2}$ under rotations in 
the $(2i-1, 2i)$ plane) and 
in an R-symmetry group $Sp(2 {\cal N})$ where ${\cal N}$ is 1 or 2. (We follow the same conventions as \cite{Bhattacharya:2008zy}.)  

We already know from our analysis of non-supersymmetric graviton scattering
that $d=6$ allows for a larger range of polarization-vectors that behave
well under the BCFW extension. In particular, if we are extending $\vect{k_1}$
and $\vect{k_n}$ in $d=4$, we are forced to take $\vect{\ep^1} = \vect{\ep^n}$.
However, we have more options in $d=6$. For example, given two vectors
$\vect{v^1}, \vect{v^2}$ that are orthogonal to each other and to $\vect{q}, \vect{k_1}, \vect{k_n}$, we can choose $\ep^1_{i j} = q_i q_j$ and 
$\ep^n_{i j} = v^{1}_{(i} v^{2}_{j)}$.  This is reflected in the fact 
that apart from the ``diagonal'' subset above, supersymmetric
theories in $d=6$ allow for
another calculable subset of correlators, which we now describe.

For simplicity, we choose a basis so that $\vect{k_1} = (1,0,0,0,0,0)$ and  $\vect{k_n} = (a,b,0,0,0,0)$ where $a$ and $b$ are arbitrary. We also define two vectors $\vect{q_1}=(0,0,0,0,1,i)$, and $\vect{q_n}=(0,0,1,i,0,0)$. We now form {\em two} arrays of  $4 {\cal N}$  supercharges each: ${\cal Q}_{1 +}^A = \{Q^I_{ \pm 1/2, \pm 1/2, 1/2} \}$,  and ${\cal Q}_{n +}^A = \{Q^I_{\pm 1/2, 1/2, \pm 1/2} \}$. 

Then, for the case of maximal supersymmetry, which is ${\cal N} = 2$, 
we make the following definitions:
\begin{equation}
\label{tnegheld6}
\begin{split}
T_{--}(\vect{k_1}) &= T_{i j}(\vect{k_1}) q_1^i q_1^j; \quad T_{--}(\vect{k_n}) = T_{i j}(\vect{k_n}) q_n^i q_n^j; \\
 U_1(\eta)&=\exp{[{\cal Q}^A_{1 +} \eta_{A}]}; \quad U_n(\eta)=\exp{[{\cal Q}^A_{n +} \eta_{A}]},
\end{split}
\end{equation}
where we have raised indices using the flat-space metric.
If we go through the procedure of listing all representations in the stress-tensor multiplet \cite{Gunaydin:1984wc} and then reducing them under the little group, we again find 
that all operators in a single multiplet are accessible by expanding the smooth functions,
\begin{equation}
\label{tcoherentd6}
T_1(\eta) = U_{1}(\eta) T_{--}(\vect{k_1}) U_{1}(-\eta); \quad T_n(\eta) = U_{n}(\eta) T_{--}(\vect{k_n}) U_{n}(-\eta); 
\end{equation}
in the Grassmann parameters $\eta^A$.

Note that we could have contracted $T_{i j}(\vect{k_n})$ with $q_1^i q_1^j$
in \eqref{tnegheld6}, but we have made a different choice above. The reason for this choice
is that we can now compute any correlator that can be written as
\begin{equation}
\label{sixcompute}
\left\langle U_1(\eta_1) U_n(\eta_n) T_{--}(\vect{k_1}) T_{--}(\vect{k_n}) O_{\mathcal C} U_n(-\eta_n) U_1(-\eta_1)\right\rangle,
\end{equation}
using BCFW recursion. This subset of correlators is significantly
larger than the ``diagonal'' subset that we described for $d=4$ above. 
This is because the expression \eqref{sixcompute} involves three-fourths of the supercharges whereas the diagonal subset just involves half. 
Moreover, we remind the reader that it is possible to make different choices (in fact, a continuous family of choices) for $\vect{q_1}$ and $\vect{q_n}$ above.\footnote{One special case is $\vect{q_1} = \vect{q_n}$, in
which case \eqref{sixcompute} reduces to a diagonal subset.} This greatly
enlarges the set of correlation functions that we can compute in maximally
supersymmetric theories in $d=6$ but it appears that this is still
not enough to compute every correlator. 

In theories with ${\cal N} = 1$ supersymmetry, the analysis above
can be repeated for multiplets that contain a conserved current. Since this
is almost identical to our analysis above, we will not repeat it explicitly.

\subsection{Supersymmetric Theories with $d = 5$}
In $d=5$, the supercharges are spinors under $SO(5)$ and the R-symmetry $SU(2)$. This algebra has a half-BPS multiplet containing a conserved current and
we can compute diagonal correlators of operators in this multiplet. This 
analysis is very similar to the analysis for $d=4$ and $d=6$ above. 

However, the stress-tensor lives in a quarter-BPS multiplet \cite{D'Auria:2000ah}. As we pointed out above, this means that not all 
operators in this multiplet can be reached via the analogue of \eqref{tcoherent}. However, as in theories with reduced supersymmetry in flat-space \cite{Lal:2009gn}, the analogue of \eqref{tcoherent} still spans a subspace of operators. 
In this subspace, we can compute diagonal correlators.

\section{Examples \label{secexamples}}
In this section, we present a few simple calculations with gauge fields to illustrate
the results that we have described above. Since the Ward identity is usually familiar
in momentum space and not in the form \eqref{wardidentity}, we start by showing how it works
in AdS. Then we verify the $w \rightarrow \infty$ behavior for a four-point gauge
amplitude. Finally, most transition amplitudes receive a divergent contribution
from the space near the boundary, which corresponds to  $z \rightarrow 0$. This needs to be regulated by cutting the space off at $z = \epsilon.$ At the end of this section, we briefly point out that this does not affect the validity 
of our recursion relations.

\subsection{Verification of the Ward Identity for a three-point function}
We will now turn on the coupling constant and examine the action for
Yang-Mills theory in \ads[5]. The action is
\begin{equation}
\label{ymaction}
S = {-1 \over 4} \int \sqrt{-g} F_{\mu \nu}^{\rm a} F^{\mu \nu, \rm{a}} d^d \vect{x} d z,
\end{equation}
where
\begin{equation}
\label{ymfieldstrength}
F_{\mu \nu}^{\rm a} = \nabla_{\mu} A_{\nu}^{\rm a} - \nabla_{\nu} A_{\mu}^{\rm a} + g_{\rm YM} f^{a b c} A_{\mu}^b A_{\nu}^c.
\end{equation}
Expanding \eqref{ymaction} using \eqref{ymfieldstrength} we see that 
perturbative Witten diagrams have a very similar structure to flat-space
perturbation theory, except that we need to replace the momenta in
flat-space Feynman diagrams with covariant derivatives. 

We will now use this to verify the Ward identity for the three-point function in Yang-Mills theory. Our object, in doing this toy calculation, is just to illustrate
the use of the Ward identity in the form \eqref{wardidentity}.
The full three-point transition amplitude is proportional to
\begin{equation}
\label{threeptexpression}
T =  \int_{\vect{x},z} a_1^{\mu} \left(a_2^{\nu} \nabla_{\mu} a_{3 \nu} - a_3^{\nu} \nabla_{\mu} a_{2 \nu} \right) + a_2^{\mu} \left( a_{3}^{\nu} \nabla_{\mu} a_{1 \nu} - a_1^{\nu} \nabla_{\mu} a_{3 \nu} \right) + a_3^{\mu} \left(a_1^{\nu} \nabla_{\mu} a_{2 \nu} - a_2^{\nu} \nabla_{\mu} a_{1 \nu} \right), 
\end{equation}
where we have suppressed an unimportant overall color-factor and a single factor of $g_{\rm YM}$ and we remind the reader that $\int_{\vect{x},z} \equiv \int d^d \, \vect{x} \, d \, z$.
Here the $\vect{a_m}$ are any solutions
to the equations of motion.
We now want to show that if we take
\begin{equation}
\label{epeqgradphi}
a_{1 \mu} = \nabla_{\mu} \phi,
\end{equation}
then this expression vanishes. We see that in this case
\begin{equation}
\label{twithgradphi}
\begin{split}
T  =   \int_{\vect{x},z} \Bigl[&\nabla^{\mu} \phi \left(a_2^{\nu} \nabla_{\mu} a_{3 \nu} - a_3^{\nu} \nabla_{\mu} a_{2 \nu} \right) + a_2^{\mu} \left( a_{3}^{\nu} \nabla_{\mu} \nabla_{\nu} \phi - \nabla^{\nu} \phi \nabla_{\mu} a_{3 \nu} \right) \\ &+ a_3^{\mu} \left(\nabla^{\nu} \phi \nabla_{\mu} a_{2 \nu} - a_2^{\nu} \nabla_{\mu} \nabla_{\nu} \phi \right) \Bigr].
\end{split}
\end{equation}
We will assume that we can integrate by parts and discard boundary terms. Below, we write equivalences up to integration by parts with a $\sim$ sign. We have 
\begin{equation}
\label{firsterm}
 \nabla^{\mu} \phi \left(a_2^{\nu} \nabla_{\mu} a_{3 \nu} - a_3^{\nu} \nabla_{\mu} a_{2 \nu} \right) \sim \phi \left(-\nabla_{\mu} a_2^{\nu} \nabla^{\mu} a_{3 \nu} - a_2^{\nu} \Box a_{3 \nu} + \nabla_{\mu} a_3^{\nu} \nabla^{\mu} a_{2 \nu} + a_{3}^{\nu} \Box a_{2 \nu} \right).
\end{equation}
The second term inside the integral \eqref{twithgradphi} is
\begin{equation}
\label{secondterm}
a_2^{\mu} a_3^{\nu} \nabla_{\mu} \nabla_{\nu} \phi - a_2^{\mu} \nabla_{\mu} a_{3 \nu} \nabla_{\nu} \phi  \sim \phi \left(\nabla_{\nu} \nabla_{\mu} (a_2^{\mu} a_3^{\nu}) + \nabla_{\nu} (a_{2}^{\mu} \nabla_{\mu} a_{3 \nu})\right).
\end{equation}
The third term is
\begin{equation}
\label{thirdterm}
a_3^{\mu} \left(\nabla^{\nu} \phi \nabla_{\mu} a_{2 \nu} - a_2^{\nu} \nabla_{\mu} \nabla_{\nu} \phi \right) \sim -\phi \left(\nabla_{\nu} \left(a_3^{\mu} \nabla_{\mu} a_{2}^{\nu} \right) + \nabla_{\nu} \nabla_{\mu} a_3^{\mu} a_2^{\nu} \right).
\end{equation}
Parts of  the second and third term cancel because
\begin{equation}
\label{secondplusthird}
\begin{split}
&\nabla_{\nu} \nabla_{\mu} a_2^{\mu} a_3^{\nu} - \nabla_{\nu} \nabla_{\mu} a_3^{\mu} a_2^{\nu} \\ &= \left(\nabla_{\nu} a_3^{\nu} \nabla_{\mu} a_2^{\mu} + a_3^{\nu} \nabla_{\nu} \nabla_{\mu} a_2^{\mu} + \nabla_{\nu} a_2^{\mu} \nabla_{\mu} a_3^{\nu} + a_2^{\mu} \nabla_{\nu} \nabla_{\mu} a_3^{\nu}\right) \\ &-  \left(\nabla_{\nu} a_3^{\nu} \nabla_{\mu} a_2^{\mu} + a_3^{\nu} \nabla_{\mu} \nabla_{\nu} a_2^{\mu} + \nabla_{\nu} a_2^{\mu} \nabla_{\mu} a_3^{\nu} + a_2^{\mu} \nabla_{\mu} \nabla_{\nu} a_3^{\nu}\right) \\
&= a_3^{\nu} \nabla_{\nu} \nabla_{\mu} a_2^{\mu} - a_3^{\nu} \nabla_{\mu} \nabla_{\nu} a_2^{\mu} +  a_2^{\mu} \nabla_{\nu} \nabla_{\mu} a_3^{\nu} - 
a_2^{\mu} \nabla_{\mu} \nabla_{\nu} a_3^{\nu} \\ &=  0.
\end{split}
\end{equation}
In the last line above, we have used the fact that both commutators of covariant derivatives lead to the Ricci tensor but with opposite signs.

The remainder of the second and third terms can be added to the first term to get something that vanishes by the equations of motion. In the first term, after a little bit of manipulation above we get
\begin{equation}
\label{vanishinganswer}
\phi \left[ \left(a_3^{\nu} \Box a_{2 \nu} - a_3^{\nu} \nabla_{\nu} \nabla_{\mu} a_2^{\mu}\right) - \left(a_2^{\nu} \Box a_{3 \nu} - a_2^{\nu} \nabla_{\nu} \nabla_{\mu} a_3^{\mu} \right) \right],
\end{equation}
which vanishes when $\vect{a_2}$ and $\vect{a_3}$ are solutions to 
the equations of motion. Note that our calculation did not involve any choice of gauge. This is important because
to derive \eqref{gaugerecurs} we needed to use the Ward identity twice, which is justified because the only assumption
in \eqref{wardidentity} is that every $A_{\mu_m}^{\rm a_m}$ is a solution to the equation of motion. This assumption
continues to hold even after we make a replacement of the form \eqref{epeqgradphi}, and so we are allowed to make
multiple such replacements.

\subsection{Verification of BCFW for a four-point function}
We now verify the BCFW recursion relations \eqref{gaugerecurs} for a four-point amplitude. 
We are really interested in the behaviour of this amplitude at $w \rightarrow \infty$ under the 
extension \eqref{bcfwextension}. What we will check here is that the integrand of this amplitude is a rational function of $w$ with no pole at $w \rightarrow \infty$. The recursion relations \eqref{gaugerecurs} then automatically
follow from the comment under \eqref{axialpropagator}. 

First,
let us understand the structure of perturbation theory. Consider, a 
four-point correlator where the four momenta are $\vect{k_1}, \ldots \vect{k_4}$. To evaluate this, we need to draw four Witten diagrams: the s,t,u channel
diagrams and a diagram involving a four-point contact interaction. Now,
the integrand of the s-channel diagram evaluates to (up to factors that we will be careful about below in the actual computation)
\[
{-1 \over 2}
\left[ a_1^{\nu} (a_2^{\mu} \lrdel_{\nu})  + a_1^{\nu} \overset{\leftrightarrow}{\nabla^{\mu}} a_{2 \nu} - a_2^{\nu} a_1^{\mu} \lrdel_{\nu} \right] G_{\mu \alpha}  \left[a_3^{\beta} \overset{\leftrightarrow}{\nabla^{\alpha}} a_{4 \beta} - (\lrdel_{\gamma} a_3^{\alpha}) a_{4}^{\gamma}  + (\lrdel_{\gamma} a_4^{\alpha}) a_{3}^{\gamma} \right],
\]
where $\vect{A} \lrdel \vect{B} \equiv \vect{A} \nabla \vect{B} - \vect{B} \nabla \vect{A}$ for two vector fields $\vect{A}$ and $\vect{B}$. $G$ is the propagator and $\vect{a_m}$ are solutions to the equations of motion. 

Now, the key point is that if we are in axial gauge, then $G$ has indices
only along the boundary directions. All covariant derivatives are covariant 
derivatives of vector fields. If we take the various $\vect{a_m}$ also to have indices only along
the Poincare directions (i.e choose axial gauge), then we can replace 
all covariant derivatives by ordinary derivatives along the boundary directions. This simplification
occurs only for the four-point amplitude. Of course, the fact that we are in
AdS shows up (a) in additional factors of $z$  and (b) in the 
propagator. But, apart from this, the expressions we obtain for the {\em integrand} essentially 
match those given by axial-gauge perturbation theory in flat-space. So, we
now start by reviewing how the BCFW extension works in flat-space and then
generalize our calculation to AdS. To further simplify our computation, 
we restrict ourselves to color-ordered amplitudes. (See \cite{Dixon:1996wi} for a review.) The full amplitude can be completely reconstructed from the color-ordered amplitudes, so we do not  lose any information by doing this. 

\subsubsection{BCFW in flat-space axial gauge:}
The rules for color-ordered amplitudes in flat-space are given in page 11 of \cite{Dixon:1996wi}. The three- and four-point vertices are\footnote{Note that the
choice of axial gauge does not affect these vertices.}
\begin{equation}
\label{threefourpt}
\begin{split}
V_3 &= {i \over \sqrt{2}} \left[ (\vect{\ep_1 } \cdot \vect{ \ep_2}) (\vect{\ep_3} \cdot \vect{(\vect{k_2}- \vect{k_1})}) + (\vect{\ep_3 } \cdot \vect{ \ep_1}) (\vect{\ep_2} \cdot \vect{(\vect{k_1}- \vect{k_3})}) + (\vect{\ep_2 } \cdot \vect{ \ep_3}) (\vect{\ep_1} \cdot \vect{(\vect{k_3}- \vect{k_2})}) \right], \\
V_4 &= i \left[(\vect{\ep_1 } \cdot \vect{ \ep_3}) (\vect{\ep_2 } \cdot \vect{ \ep_4}) - {1 \over 2} (\vect{\ep_1 } \cdot \vect{ \ep_2}) (\vect{\ep_3 } \cdot \vect{ \ep_4}) - {1 \over 2} (\vect{\ep_1 } \cdot \vect{ \ep_4}) (\vect{\ep_2 } \cdot \vect{ \ep_3})  \right],
\end{split}
\end{equation}
where the $\vect{\ep_n}$ are now polarization vectors.\footnote{\label{flatvsads} As we pointed out in section \ref{sectransitionamp}, in flat-space perturbation theory we 
only need to make reference to the polarization vectors, whereas in AdS we need
to consider the solution to the equation of motion associated to a
given polarization vector because this has a nontrivial dependence on $z$.}
There are three diagrams that contribute to the four-point correlator. The four-point vertex, the s-channel diagram where $1$ and $2$ meet at a point and the
$t$ channel diagram where $1$ and $4$ meet at a point. We are dealing only with
color-ordered correlators, so we do not need to worry about the $u$-channel.

These three terms are given by
\begin{equation}
\label{t1term}
T_1 = i \left[(\vect{\ep_1 } \cdot \vect{ \ep_3}) (\vect{\ep_2 } \cdot \vect{ \ep_4}) - {1 \over 2} (\vect{\ep_1 } \cdot \vect{ \ep_2}) (\vect{\ep_3 } \cdot \vect{ \ep_4}) - {1 \over 2} (\vect{\ep_1 } \cdot \vect{ \ep_4}) (\vect{\ep_2 } \cdot \vect{ \ep_3})  \right],
\end{equation}
which comes from the four-point vertex. The s-channel diagram gives
\begin{equation}
\label{t2term}
\begin{split}
T_2 = &{i \over 2 \left[(\vect{k_1}+ \vect{k_2})^2 + p^2 \right]} \left[ \left\{ (\vect{\ep_1} \cdot \vect{\ep_2}) (\vect{k_2}- \vect{k_1})^i + (2 \vect{\ep_2} \cdot \vect{k_1}) \ep_1^i - 2 (\vect{\ep_1 } \cdot \vect{ k_2}) \ep_2^i \right\} \right. \\ 
&\times  \left( \eta_{i j} + {K_i K_j \over p^2} \right) \left. \left\{(\vect{\ep_3 } \cdot \vect{ \ep_4}) (\vect{k_4}- \vect{k_3})^j + 2 (\vect{\ep_4} \cdot \vect{k_3} ) \ep_3^j  - 2(\vect{\ep_3 } \cdot \vect{ k_4}) \ep_4^j \right\}\right],
\end{split}
\end{equation}
where $\vect{K} = -(\vect{k_1} + \vect{k_2}) = (\vect{k_3} + \vect{k_4})$ and we
have chosen the notation $p = K^0$ to indicate the  analogy with AdS.  Note that when the terms in brackets
are dotted with the part of the propagator that contains $\vect{K}$, we get some simplifications because we get 
\begin{equation}
\label{simplificationinfourpt}
(\vect{\ep_2 } \cdot \vect{ k_1}) (\vect{\ep_1 } \cdot \vect{K}) - (\vect{\ep_1 } \cdot \vect{ k_2}) (\vect{\ep_2 } \cdot \vect{K}) = (\vect{\ep_2 } \cdot \vect{ k_1}) (\vect{\ep_1 } \cdot \vect{ k_2}) - (\vect{\ep_1 } \cdot \vect{ k_2}) (\vect{\ep_2 } \cdot \vect{ k_1}) = 0.
\end{equation}
This leads to
\begin{equation}
\label{t2simple}
\begin{split}
T_2 = {i \over 2 \left[(\vect{k_1}+ \vect{k_2})^2 + p^2 \right]} &\left[ \left\{ (\vect{\ep_1} \cdot \vect{\ep_2}) (\vect{k_2}- \vect{k_1})^i + (2 \vect{\ep_2} \cdot \vect{k_1}) \ep_1^i - 2 (\vect{\ep_1 } \cdot \vect{ k_2}) \ep_2^i \right\} \eta_{i j} \right. \\ 
&\times \left. \left\{(\vect{\ep_3} \cdot \vect{ \ep_4}) (\vect{k_4}- \vect{k_3})^j + 2 (\vect{\ep_4} \cdot \vect{k_3}) \ep_3^j  - 2(\vect{\ep_3 } \cdot \vect{ k_4}) \ep_4^j \right\} \right. \\ &\left. + (\vect{\ep_1 } \cdot \vect{ \ep_2}) (\vect{\ep_3 } \cdot \vect{ \ep_4}) {(\vect{k_1}^2 - \vect{k_2}^2) (\vect{k_4}^2 - \vect{k_3}^2) \over p^2} \right]. 
\end{split}
\end{equation}
The third term is just obtained by left-shifting $(1,2,3,4) \rightarrow (4,1,2,3)$.  This gives us
\begin{equation}
\label{t3term}
\begin{split}
T_3 = {i \over 2 \left[(\vect{k_4}+ \vect{k_1})^2 + p^2 \right]} &\left[ \left\{ (\vect{\ep_4} \cdot \vect{\ep_1}) (\vect{k_1}- \vect{k_4})^i + (2 \vect{\ep_1} \cdot \vect{k_4}) \ep_4^i - 2 (\vect{\ep_4 } \cdot \vect{ k_1}) \ep_1^i \right\} \eta_{i j} \right. \\ 
&\times \left. \left\{(\vect{\ep_2 } \cdot \vect{ \ep_3}) (\vect{k_3}- \vect{k_2})^j + 2 (\vect{\ep_3} \cdot \vect{k_2}) \ep_2^j  - 2(\vect{\ep_2 } \cdot \vect{ k_3}) \ep_3^j \right\} \right. \\ &\left. + (\vect{\ep_4 } \cdot \vect{ \ep_1}) (\vect{\ep_2 } \cdot \vect{ \ep_3}) {(\vect{k_4}^2 - \vect{k_1}^2) (\vect{k_3}^2 - \vect{k_2}^2) \over p^2} \right].
\end{split}
\end{equation}
The full answer is 
\begin{equation}
\label{texp}
T = T_1 + T_2 + T_3.
\end{equation}
We now analyze this with some specific choices for $\vect{\ep_1}$ and $\vect{\ep_4}$.

\paragraph{$\vect{\ep_1} = \vect{\ep_4} = \vect{q}$:}
We start with this case because this is
the case we are most interested in for AdS.
Let us take $\vect{k_1}\rightarrow \vect{k_1}+ \vect{q} w, \vect{k_4}\rightarrow \vect{k_4}- \vect{q} w$ and then
consider the behaviour of the four-point amplitude. Naively, it would
seem that we have $O(1)$ terms but we would like all of them
to cancel. (The $O(w)$ terms cancel automatically here.)

Note that since $\vect{q} \cdot \vect{k_1}= \vect{q} \cdot \vect{k_4}= 0$, we have $\vect{q} \cdot \vect{k_2}= - \vect{q} \cdot \vect{k_3}$. Also, with this choice of polarization vectors $T_3 = 0$. 
The $\Or[1]$ terms in $T_2$ are 
\begin{equation}
\label{t2case1o1}
\begin{split}
\lim_{w \rightarrow \infty} T_2 &= {i w \over 2 \times  2 (\vect{q } \cdot \vect{ k_2}) w} (\vect{q } \cdot \vect{ \ep_2}) (\vect{q } \cdot \vect{ \ep_3}) (\vect{q } \cdot \vect{ k_2}) w \left(-2 + 2 + 2 + 2 -4 +2 -4 \right)\\
&= {-i \over 2} (\vect{q} \cdot \ep_2) (\vect{q } \cdot \vect{ \ep_3}).
\end{split}
\end{equation}
Also, we have
\begin{equation}
\label{t1case1o1}
\lim_{w \rightarrow \infty} T_1 = {i \over 2} (\vect{q } \cdot \vect{ \ep_3}) (\vect{q } \cdot \vect{ \ep_2}), 
\end{equation}
so that 
\begin{equation}
\label{case1ans}
\lim_{w \rightarrow \infty} (T_1 + T_2 + T_3) = 0.
\end{equation}

\paragraph{$\vect{\ep_1} = \vect{q}, \text{~and~} \vect{\ep_4} \neq \vect{q}, \text{~and~} \vect{\ep_4} \cdot \vect{k_1} = \vect{\ep_4} \cdot \vect{q} = 0$:}
This is the other case that is admissible for gauge fields in AdS. In this 
case, we see that the expression for $T_1$ becomes
\begin{equation}
\label{t1choice2}
T_1 = i \bigl[(\vect{q} \cdot \vect{\ep_3}) (\vect{\ep_2} \cdot \vect{\ep_4}) - {1 \over 2} (\vect{q} \cdot \vect{\ep_2}) (\vect{\ep_3} \cdot \vect{\ep_4}) \bigr].
\end{equation}
The expression for $T_2$ becomes 
\begin{equation}
\label{t2choice2}
\begin{split}
T_2 = &{i \over 4 \vect{k_2} \cdot \vect{q}} \Bigl[ \bigl\{ (\vect{q} \cdot \vect{\ep_2}) (\vect{\ep_3} \cdot \vect{\ep_4}) (\vect{q} \cdot \vect{k_3}) - 2 (\vect{q} \cdot \vect{\ep_2}) (\vect{\ep_4} \cdot \vect{k_3}) (\vect{q} \cdot \vect{\ep_3}) - 2(\vect{\ep_2} \cdot \vect{q}) (\vect{\ep_3} \cdot \vect{\ep_4}) (\vect{q} \cdot \vect{k_3}) \\ &+ 4 (\vect{\ep_2} \cdot \vect{q}) (\vect{\ep_4} \cdot \vect{k_3}) (\vect{q} \cdot \vect{\ep_3})  \bigr\} + \bigl\{2 (\vect{\ep_3} \cdot \vect{\ep_4}) (\vect{q} \cdot \vect{k_2}) (\vect{q} \cdot \vect{\ep_2}) - (\vect{\ep_3} \cdot \vect{\ep_4}) (\vect{q} \cdot \vect{\ep_2}) (\vect{q} \cdot \vect{k_2}) \\
&+ 2 (\vect{q} \cdot \vect{\ep_3}) (\vect{q} \cdot \vect{\ep_2}) \bigl((\vect{\ep_4} \cdot \vect{k_2}) - (\vect{\ep_4} \cdot \vect{k_1})\bigr) - 4 (\vect{q} \cdot \vect{\ep_3}) (\vect{q} \cdot \vect{k_2}) (\vect{\ep_4} \cdot \vect{\ep_2}) \bigr\} \Bigr] + \Or[{1 \over w}]
\end{split}
\end{equation}
This  simplifies to
\begin{equation}
\label{t1choice2simple}
\begin{split}
T_1 = {-i \over  \vect{k_2} \cdot \vect{q}} &\Bigl[ (\vect{q} \cdot \vect{\ep_2}) (\vect{q} \cdot \vect{\ep_3}) (\vect{k_1} \cdot \vect{\ep_4})  +  (\vect{q} \cdot \vect{\ep_3}) (\vect{q} \cdot \vect{k_2})(\vect{\ep_4} \cdot \vect{\ep_2}) - {1 \over 2} (\vect{\ep_3} \cdot \vect{\ep_4}) (\vect{q} \cdot \vect{k_2}) (\vect{q} \cdot \vect{\ep_2})\Bigr] \\ &+ \Or[{1 \over w}]\\
= {-i \over  \vect{k_2} \cdot \vect{q}} &\Bigl[ (\vect{q} \cdot \vect{\ep_2}) (\vect{q} \cdot \vect{\ep_3}) (\vect{k_1} \cdot \vect{\ep_4})\Bigr]  - i (\vect{q} \cdot \vect{\ep_3}) (\vect{\ep_4} \cdot \vect{\ep_2}) + {i \over 2} (\vect{\ep_3} \cdot \vect{\ep_4}) (\vect{q} \cdot \vect{\ep_2}) + \Or[{1 \over w}] 
\end{split}
\end{equation}
There are also $\Or[1]$ terms in $T_3$. These are given by
\begin{equation}
\label{t3choice2}
T_3 = {-2 i (\vect{\ep_4} \cdot \vect{k_1}) \over 2 (\vect{k_1} + \vect{k_4})^2}   \Bigl[ (2 \vect{q} \cdot \vect{k_3}) (\vect{\ep_2} \cdot \vect{\ep_3}) 
+ (\vect{q} \cdot \vect{\ep_2}) (2 \vect{\ep_3} \cdot \vect{k_2}) - 2 (\vect{\ep_2} \cdot \vect{k_3}) (\vect{q} \cdot \vect{\ep_3}) \Bigr] + \Or[{1 \over w}].
\end{equation}
With $\vect{\ep_4} \cdot \vect{k_1} = 0,$ we see that once again
\begin{equation}
\label{noboundarywinf}
\lim_{w \rightarrow \infty} (T_1 + T_2 + T_3) = 0.
\end{equation}
We now turn to the analysis in AdS.

\subsubsection{BCFW in AdS}
We will now demonstrate that with a simple trick, the calculation above
can be generalized to show that the integrand of the AdS transition amplitude
behaves well under the BCFW extension. When we generalize to the computation
of the AdS transition amplitude, the interaction 
vertices change only in that we get factors of $z$
when we dot vectors into one another and when we raise vectors. Say, 
we start with all vectors --- $\vect{a_m}$ and $\vect{k_m}$ --- lowered. The $\vect{a_m}$ are, as above, solutions to the equations of motion with polarization
vectors $\vect{\ep_m}$. Then, while keeping
the notation $b \cdot c = b_i c_j \eta^{i j}$ and writing factors of $z$
explicitly, we can adapt the result \eqref{t2simple} to AdS:
\begin{equation}
\label{t2ads}
\begin{split}
T_2 = \int &{d z \over z^{d + 1}} {d z' \over (z')^{d + 1}} {d p^2 \over 2} {i \over 2 \left[(\vect{k_1}+ \vect{k_2})^2 + p^2 \right]} z^4 (z')^4 z^{\nu_1} (z')^{\nu_1} J_{\nu_1} (p z) J_{\nu_1} (p z')  \\ 
&\times \Bigl[ \bigl\{ (\vect{a_1} \cdot \vect{a_2}) (\vect{k_2}- \vect{k_1})^i + (2 \vect{a_2} \cdot \vect{k_1}) a_1^i - 2 (\vect{a_1 } \cdot \vect{ k_2}) a_2^i \bigr\} \eta_{i j}  \\ 
&\phantom{\times \Bigl[}\times \bigl\{(\vect{a_3 } \cdot \vect{ a_4}) (\vect{k_4}- \vect{k_3})^j + 2 (\vect{a_4} \cdot \vect{k_3}) a_3^j  - 2(\vect{a_3 } \cdot \vect{ k_4}) a_4^j \bigr\}  \\ 
&\phantom{\times \Bigl[}  + (\vect{a_1 } \cdot \vect{ a_2}) (\vect{a_3 } \cdot \vect{ a_4}) {(\vect{k_1}^2 - \vect{k_2}^2) (\vect{k_4}^2 - \vect{k_3}^2) \over p^2} \Bigr],\\
\end{split}
\end{equation}
Here we have performed the integral over the boundary directions and imposed
momentum conservation at each vertex.   We also remind the reader that $\nu_1 = {d - 2 \over 2}$. The $\vect{a_m}$ remain functions of the
radial coordinate. It is important to note that $\vect{a_1}$ and $\vect{a_2}$ are functions of $z$ while $\vect{a_3}$ and $\vect{a_4}$ are functions of $z'$
although we have not shown this explicitly in  \eqref{t2ads} to lighten
the notation.

On the other hand,
\begin{equation}
\label{t1ads}
T_1 = \int {i d z \over z^{d + 1}} z^4  \left[(\vect{a_1 } \cdot \vect{ a_3}) (\vect{a_2 } \cdot \vect{ a_4}) - {1 \over 2} (\vect{a_1 } \cdot \vect{ a_2}) (\vect{a_3 } \cdot \vect{ a_4}) - {1 \over 2} (\vect{a_1 } \cdot \vect{ a_4}) (\vect{a_2 } \cdot \vect{ a_3})  \right].
\end{equation}
At first sight it looks like under the BCFW extension the expression \eqref{t1ads}, which
is an integral over a single spacetime point, and the expression \eqref{t2ads}, which 
involves an integral over two spacetime points, will behave very differently.
 The trick is to {\em split} the single point in \eqref{t1ads} by using the Bessel function closure relation:
\begin{equation}
\label{hankel2}
\int J_{\nu_1} (p z) J_{\nu_1} (p z')  {d p^2 \over 2}  = {\delta(z - z') \over z}.
\end{equation}
This allows us to write
\begin{equation}
\label{t1splitpoint}
\begin{split}
T_1 = i \int &{ d z \over z^{d + 1}} {d z' \over (z')^{d + 1}} {d p^2 \over 2}  z^4 (z')^4 \\ &\times \Bigl[(\vect{a_1}(z) \cdot \vect{a_3}(z')) (\vect{a_2}(z) \cdot \vect{a_4}(z')) - {1 \over 2} (\vect{a_1}(z) \cdot \vect{a_2}(z)) (\vect{a_3}(z') \cdot \vect{a_4}(z'))  \\ 
&\phantom{\times \Bigl[}
- {1 \over 2} (\vect{a_1}(z) \cdot \vect{a_4}(z')) (\vect{a_2}(z) \cdot \vect{a_3}(z'))  \Bigr] z^{\nu_1} J_{\nu_1} (p z) (z')^{\nu_1} J_{\nu_1} (p z').
\end{split}
\end{equation}
Note that the factors of $z$ work out correctly. When we integrate over $p$
in \eqref{t1splitpoint}, we set $z = z'$ and get an extra factor of ${1 \over z}$. So the total power of $z$ works out to $8 + 2 \nu_1  - 2 (d+1) - 1 = 4 - (d+1)$,
which is the same as \eqref{t1ads}. 

In this form, we can repeat the calculations above for both choices of
external polarization vectors. Under the BCFW extension \eqref{bcfwextension},
 it is clear that the $\Or[1]$ term in the integrand cancels between $T_1$ and $T_2$. Note that we do not need to worry about $T_3$ since its $\Or[1]$ contribution vanishes in AdS also. The recursion relations \eqref{gaugerecurs} now follow from Cauchy's 
residue theorem for the integrand.

\subsection{Divergences from the Boundary}
This is a good place to pause and examine divergences from the boundary. 
Position-space CFT correlators
have short-distance divergences when two points come close to each other. When we transform
these correlators into momentum space, we integrate over all positions; we cannot exclude the configurations where two points coincide. 
In the momentum-space transition amplitudes that we have been considering, the short-distance
singularities of position-space correlators manifest
themselves in divergent contributions from the part of the integral near
the boundary. To regulate these divergences, we need to cut off the space at 
some finite value $z = \epsilon,$ and then discard away the terms that, when 
Fourier transformed, lead to short-distance singularities.

This does not affect our recursion relations, which just rely on the fact that the {\em integrand} is a rational function of the momenta, which can be recovered from its residues. So, these recursion relations do not see the limits
of the radial-integrals. This is clear in the example above where we never needed
to know the range of $z$ or $z'$.\footnote{We emphasize that we did need to fix
the limits of the $p$-integral in \eqref{hankel2}. We also required the
fact that $z$ and $z'$ vary over the {\em same} range.} Hence, if we allow
the $z$ and $z'$ integrals to run from $(\epsilon, \infty)$ instead of $(0, \infty)$,
our computation will be unaffected.

\section{Results and Discussion \label{secresults}}
The key objects of study in this paper were transition amplitudes,
which are defined precisely in \eqref{transdef}. These include vacuum
correlators of the boundary theory as a special case. We showed that these 
transition amplitudes obey the recursion relations \eqref{gaugerecurs} in
conformal field theories with a bulk Yang-Mills dual and the recursion
relations \eqref{gravityrecurs} in CFTs with a bulk gravity dual. For interacting bulk scalars, we need to add an additional boundary term
that is shown explicitly in  \eqref{scalarecurs}. These
recursion relations reproduce the results of tree-level Witten diagrams
but are more efficient. 

The conditions on polarization vectors, for a transition amplitude to be
well-behaved under the BCFW extension, are stronger than in flat-space. 
For Yang-Mills theory these are enumerated in section \ref{gaugebossec}  and for gravity 
they are given in section \ref{gravitysec}. We showed that for a bulk Yang-Mills
theory even without using constraints imposed by conformal
symmetry, any arbitrary configuration 
of external polarization vectors could be built up by combining different
BCFW extensions. This is also true for gravity in $d=6$ and higher. For gravity in $d=4$,
we can calculate 624 out of 625 possible polarization-combinations for a
four-point function and all possible polarizations for five- and higher-point functions. For gravity in $d=5$, we need at least six external particles before
we can access all possible polarization-combinations.

In section \ref{secsusytheories}, we generalized these recursion relations
to theories with supersymmetry.  Supersymmetry allows us to compute additional correlators where
we can convert at least two operators to conserved currents or stress tensors
with appropriate polarizations. However, the stronger constraints on the
polarization-combinations, which are well behaved under a BCFW extension, implies that not all correlators of operators
in the same supersymmetry multiplet as the stress-tensor or a conserved
current are calculable by these techniques. In particular, for 
operators in the same multiplet as the stress-tensor in $d=4$, 
maximal supersymmetry allows us to compute
 the ``diagonal'' subset of operators \eqref{alltostress}. In $d=6$,
a larger subset of operators is accessible: apart from the diagonal
subset we can also calculate operators of the form \eqref{sixcompute}.

There are several directions in which this investigation can be extended.\footnote{One line 
of inquiry, which is somewhat orthogonal to the perspective here, but quite interesting,
is to explore whether an analogue of these recursion relations can be written for Mellin-transformed position-space correlators \cite{Penedones:2010ue,Mack:2009mi,Mack:2009gy}.} In flat-space, the BCFW recursion relations turn out to be surprisingly 
useful at one and higher loops. It would be nice to generalize this to
AdS. This would incorporate ${1 \over N}$ corrections in the bulk. On the other hand,
it would also be interesting to try and incorporate $\alpha'$ corrections.
A  version of the BCFW recursion relations also seems to work
for flat-space string theory \cite{Boels:2008fc,Boels:2010bv,Cheung:2010vn}. What about string theory in AdS? Is it at least possible to extend these recursion relations to simple
nonlocal theories, as one can do with noncommutative theories \cite{Raju:2009yx} in 
flat-space?

In another direction, it would be interesting to understand if there is 
an analogue of the ``twistor-transform'' that allowed the authors of \cite{ArkaniHamed:2009si}
to write down a simple equation for the generating function of 
scattering amplitudes. Particularly, if we could make precise the intuition of
section \ref{sectransitionamp} and write transition amplitudes as correlators
in global AdS, we would get a simple equation for the 
generating function of stress-tensor operators on the boundary. In some sense,
this would be a ``master-field'' equation for strongly coupled ${\cn=4}$ SYM. Yet
another interesting question in this direction is whether we can use these
recursion relations to restrict the possible set of conformal field theories 
that have gravity duals \cite{ElShowk:2011ag}.  There are several other interesting properties of scattering amplitudes in flat
space such as the Kawai-Lewellen-Tye relations between gauge and gravity
amplitudes \cite{Kawai:1985xq}. Do these hold, albeit in a modified form,  in AdS?  
 
Finally, the physical intuition presented in this paper suggests that these
techniques would go through in the presence of a black-hole in the bulk. This
would now correspond to stress-tensor correlators computed at finite-temperature
on the boundary. The two-point function for the stress tensor calculated for 
thermal ${\cn=4}$ SYM in this manner \cite{Policastro:2001yc} has been quite important for investigations
at the Relativistic Heavy Ion Collider. It would be fascinating to explore whether four-
and higher-point correlators also have phenomenological implications for 
heavy-ion physics and in other systems.

\section*{Acknowledgments:} This idea, of considering the BCFW recursion
relations in AdS, first originated in a discussion with Shiraz Minwalla in April 2009. I am also grateful to Rajesh Gopakumar, Joao Penedones and  Ashoke Sen for discussions. I 
would like to acknowledge the support of a
Ramanujan fellowship and the Harvard University Physics Department.

\bibliographystyle{JHEP-2}
\bibliography{references}
\end{document}